\documentclass[a4paper,11pt]{article}
\pdfoutput=1
\usepackage[colorlinks=true]{hyperref}
\usepackage{graphicx,rotating,hyperref,slashed,amsmath,xcolor,amssymb,amsfonts,colortbl,cite}
\makeatletter
\hypersetup{colorlinks,bookmarksopen,bookmarksnumbered,  
linkcolor=blus,pdfstartview=FitH,urlcolor=rossos,citecolor=verde}
\allowdisplaybreaks		 

\hypersetup{%
    ,urlcolor=black
    ,citecolor=black
    ,linkcolor=black
    }
\usepackage{mciteplus}

\AtBeginDocument{
  \hypersetup{
    urlcolor=blue,
    citecolor=blue,
    linkcolor=black,
  }%
}

\numberwithin{equation}{section} 
 
\def\beq{\begin{equation}}
\def\eeq{\end{equation}}
 
\def\bea{\begin{eqnarray}} 
\def\eea{\end{eqnarray}}

\def\0{{\boldsymbol 0}}

\def\lsim{\mathrel{\rlap{\lower3pt\hbox{\hskip0pt$\sim$}}
   \raise1pt\hbox{$<$}}}         
\def\gsim{\mathrel{\rlap{\lower4pt\hbox{\hskip1pt$\sim$}}
   \raise1pt\hbox{$>$}}}         

 \newcommand{\sfootnote}[1]{}
\definecolor{bluc}{cmyk}{1,1,0,0.1}
\definecolor{rossoCP3}{cmyk}{0,.88,.77,.40}
\definecolor{rosso}{cmyk}{0,1,1,0.4}
\definecolor{rossos}{cmyk}{0,1,1,0.55}
\definecolor{rossoc}{cmyk}{0,1,1,0.2}
\definecolor{verdes}{cmyk}{0.92,0,0.59,0.4}

\hypersetup{colorlinks, bookmarksopen, bookmarksnumbered,
citecolor=verdes, linkcolor=bluc, pdfstartview=FitH, urlcolor=rossos}

\newcommand{\mio}[1]{}

\definecolor{Gray}{gray}{0.95}

\usepackage{multicol}
\usepackage{color}
\definecolor{rosso}{cmyk}{0,1,1,0.4}
\definecolor{rossos}{cmyk}{0,1,1,0.55}
\definecolor{rossoc}{cmyk}{0,1,1,0.2}
\definecolor{blu}{cmyk}{1,1,0,0.3}
\definecolor{blus}{cmyk}{1,1,0,0.6}
\definecolor{bluc}{cmyk}{1,1,0,0.1}
\definecolor{verde}{cmyk}{0.92,0,0.59,0.25}
\definecolor{verdec}{cmyk}{0.92,0,0.59,0.15}
\definecolor{verdes}{cmyk}{0.92,0,0.59,0.4}

\skip\@mpfootins = \skip\footins
\def\circa#1{\,\raise.3ex\hbox{$#1$\kern-.75em\lower1ex\hbox{$\sim$}}\,}

\newcommand{\be}{\begin{equation}}
\newcommand{\ee}{\end{equation}}
\newfam\rsfsfam
\def\mathscr#1{{\fam\rsfsfam\relax#1}}

\def\circa#1{\,\raise.3ex\hbox{$#1$\kern-.75em\lower1ex\hbox{$\sim$}}\,}
\makeatletter

\def\hhref#1{\href{http://arxiv.org/abs/#1}{arXiv:#1}} 

\newcommand{\doi}[1]{\href{http://dx.doi.org/#1}{[doi]}}

\def\hhref#1{\href{http://arxiv.org/abs/#1}{arXiv:#1}} 
 
\def\art{\@ifnextchar[{\eart}{\oart}}
\def\eart[#1]#2#3#4#5#6{{\rm #2}, {\em #3 \bf #4} {\rm (#6) #5} ({\em #1})}

\def\article{\@ifnextchar[{\earticle}{\oarticle}}
\def\oarticle#1#2#3#4#5#6{{\rm #1}, {\em ``#6''}, {\rm #2 #3 (#5) #4}}
\def\earticle[#1]#2#3#4#5#6#7{{\rm #2}, {\em ``#7''}, {\rm #3 #4 (#6) #5}  [\hhref{#1}]}
\def\hepart[#1]#2{{\rm #2, \em#1}}
\def\heparticle[#1]#2#3{#2, {\em ``#3''} [\hhref{#1}]}

\newcounter{alphaequation}[equation]
\def\thealphaequation{\theequation\hbox to
0.6em{\hfil\alph{alphaequation}\hfil}}
\def\eqnsystem#1{
\def\@eqnnum{{\rm (\thealphaequation)}}
\def\@@eqncr{\let\@tempa\relax \ifcase\@eqcnt \def\@tempa{& & &} \or
  \def\@tempa{& &}\or \def\@tempa{&}\fi\@tempa
  \if@eqnsw\@eqnnum\refstepcounter{alphaequation}\fi
\global\@eqnswtrue\global\@eqcnt=0\cr}
\refstepcounter{equation} \let\@currentlabel\theequation \def\@tempb{#1}
\ifx\@tempb\empty\else\label{#1}\fi
\refstepcounter{alphaequation}
\let\@currentlabel\thealphaequation
\global\@eqnswtrue\global\@eqcnt=0 \tabskip\@centering\let\\=\@eqncr
$$\halign to \displaywidth\bgroup \@eqnsel\hskip\@centering
$\displaystyle\tabskip\z@{##}$&\global\@eqcnt\@ne
\hskip2\arraycolsep\hfil${##}$\hfil& \global\@eqcnt\tw@\hskip2\arraycolsep
$\displaystyle\tabskip\z@{##}$\hfil
\tabskip\@centering&\llap{##}\tabskip\z@\cr}
\def\endeqnsystem{\@@eqncr\egroup$$\global\@ignoretrue} \makeatother

\definecolor{fiorentina}{rgb}{.5,0,.5}

\begin{document}

\setcounter{page}{1} \baselineskip=15.5pt \thispagestyle{empty}

\bigskip\

\vspace{1cm}
\begin{center}

{\fontsize{19}{28}\selectfont  \sffamily \bfseries 
{
An analytic approach to  non-slow-roll inflation
}}

\end{center}

\vspace{0.2cm}

\begin{center}
{\fontsize{13}{30}\selectfont  Gianmassimo Tasinato } 
\end{center}

\begin{center}

\vskip 8pt
\textsl{ Physics Department, Swansea University, SA28PP, UK }\\
\vskip 7pt

\end{center}

\smallskip
\begin{abstract}
\noindent
Brief periods of non-slow-roll evolution during inflation can produce interesting
observable consequences, as   primordial black holes,
or an inflationary gravitational wave spectrum enhanced  at small scales.  We develop a model independent, analytic approach for studying
  the predictions of single-field scenarios which include short phases of slow-roll violation. Our method
  is based on Taylor expanding  the equations for cosmological fluctuations in 
  a small quantity, which parameterizes the duration of the non-slow-roll eras. 
   The  super-horizon spectrum  of perturbations
  is described by few effective parameters, and is characterized by a pronounced dip followed
  by a rapid growth in its amplitude, as  typically found  
   in numerical and  analytical studies. 
   The dip position $k_{\rm dip}/k_*$
    and  the maximal enhancement $\Pi_{\rm max}$ of the spectrum towards small scales are found to be related by
    the  law $k_{\rm dip}/k_*\propto 
    \Pi_{\rm max}^{-1/4}
    $, and we determine the proportionality constant.
     For a single epoch of slow-roll violation 
   we confirm previous studies, finding that  the steepest
   slope of the spectrum well after the dip has spectral index  $n-1\,=\,4$.   On the other hand,   with   multiple phases of slow-roll violation,  the   slope of the spectrum 
      is  generally enhanced. For example, when two epochs  of slow-roll violation take place, separated by a phase of quasi-de Sitter expansion,  we find that 
   the  spectral index  can reach the value $n-1\,=\,8$. This phenomenon indicates that 
   the  slope of the spectrum keeps memory  of the  history 
   of non-slow-roll phases occurred during inflation. 
\end{abstract}

\section{Introduction   }

Single-field slow-roll inflation has many realizations, but leads  to  general predictions that
are valid for all models,  as for example  Maldacena consistency relations 
\cite{Maldacena:2002vr} between  certain limits of  inflationary correlation functions.
  So far, many of single-field   inflation predictions  have been tested  by CMB and large-scale structure 
   observations   at the largest cosmological scales.
   It is
  interesting to explore inflationary scenarios including 
  brief periods of  non-slow-roll evolution, which
   leave distinct imprints in the  small scale range of the spectrum of scalar and tensor fluctuations. We ask
   whether these scenarios lead to generic predictions valid for all models of single-field inflation. There are various
   motivations to consider these set-up.  
  A strong amplification  of the spectrum of  
   scalar curvature fluctuations could   overcome the threshold of 
 formation for  primordial black holes \cite{Hawking:1971ei,Carr:1974nx} within single-field  models of inflation, see e.g. \cite{Garcia-Bellido:2017mdw,Ezquiaga:2017fvi,Ballesteros:2017fsr,Hertzberg:2017dkh,Motohashi:2017kbs,Cicoli:2018asa,Ozsoy:2018flq} (see also \cite{Carr:2016drx,Sasaki:2018dmp,Carr:2020xqk} for reviews). Enhancing the primordial gravitational wave spectrum towards large frequencies would make it 
   detectable  with gravitational wave experiments (see e.g. \cite{Bartolo:2016ami} for a recent study focussed
 on LISA).   
 Any observation of these phenomena would be crucial for our understanding of the physics driving
 inflation, and it is then important to have a reliable theoretical understanding of the physics of inflation when brief 
 non-slow-roll phases take place. 
 \begin{figure}[h!]
\centering
 \includegraphics[width = 0.6 \textwidth]{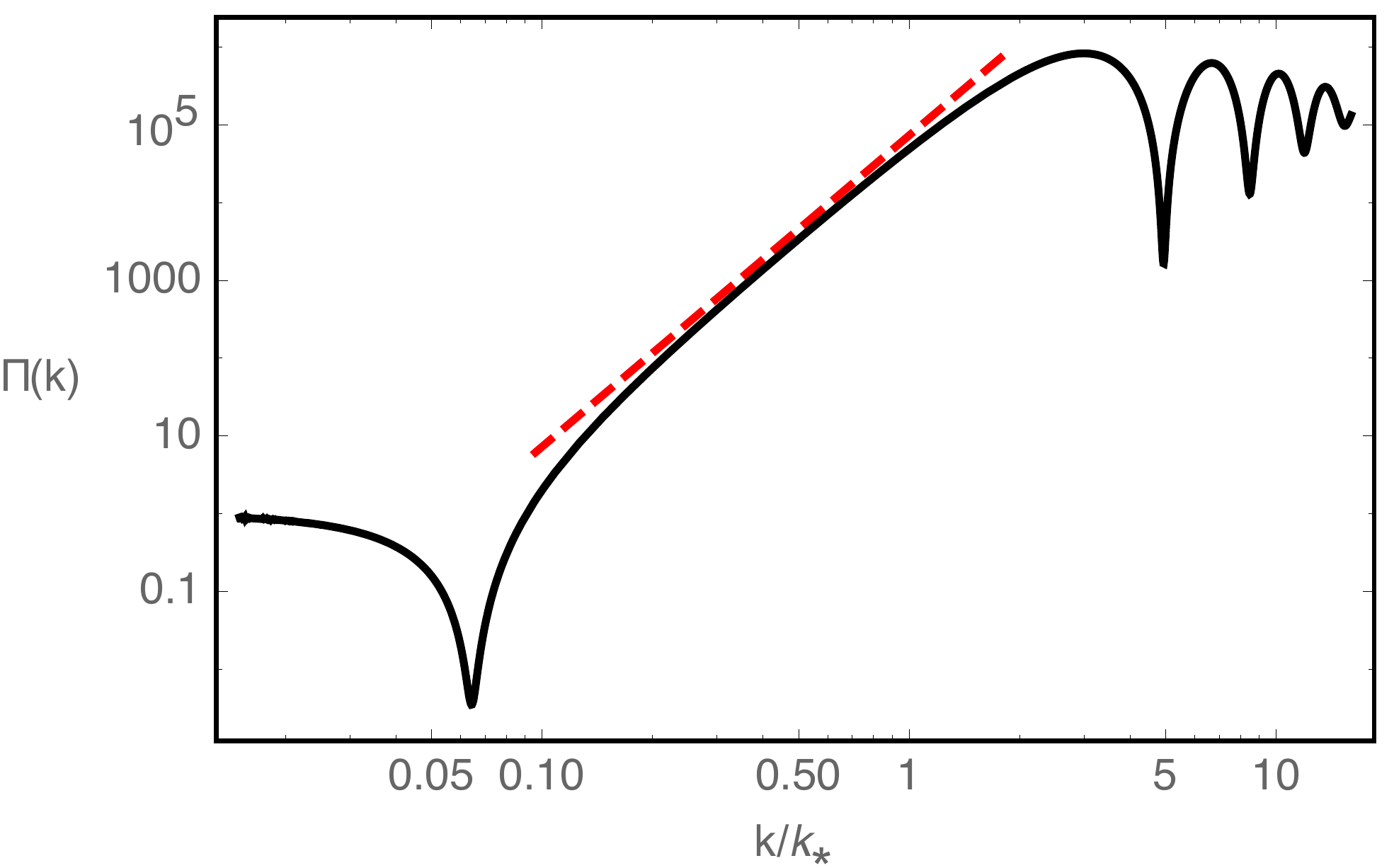}
 \caption{\it Representative plot of  the super-horizon  spectrum  of fluctuations in models with transient
 violations of slow-roll conditions  during inflation. The horizontal axis contains the ratio  $k/k_\star$, with
  $k_\star$ being the horizon-crossing scale 
  for modes leaving the horizon at the onset of the non-slow-roll epoch.  The vertical axis contains the ratio $\Pi(k)$ of the  spectrum evaluated at scale $k$, versus the large-scale spectrum at $k=0$. The red dashed line 
  has a profile proportional to $k^4$. }
 \label{fig:complete-profile2-intro}
\end{figure}

 Fig \ref{fig:complete-profile2-intro} represents 
a typical plot of the spectrum of scalar fluctuations at superhorizon scales in Fourier space, associated with single-field   models with a short non-slow-roll epoch. It contains interesting features that we would like 
to examine  in this work:
\begin{itemize}
\item[1.] The spectrum has a pronounced dip at relatively large scales, much larger than the scale at which modes
start to leave the horizon during the non-slow-roll era.  Around the dip  the spectrum profile starts to
 strongly deviate from the single-field slow-roll predictions. It is  then important  
  to better analyse the properties of the dip, as its position and depth, which would be 
   relevant for CMB and LSS constraints.
  \item[2.] After the dip, the spectrum starts to steadily increase towards its peak, with a slope that
   can not overcome certain values. For example, it has been shown \cite{Byrnes:2018txb} that  
   in single-field inflation with a single brief phase of non-attractor evolution the spectral index is 
   bounded 
  by  $n-1\,=\,4$.  It would be interesting
  to develop new analytical tools    for better understanding this bound in general models that violate slow-roll conditions.
  \item[3.]  In fact, the bound on the slope of the spectrum discussed in the previous point can to be  violated
  in models with  additional phases of non-slow-roll evolution, preceding the non-attractor era during which 
most of the spectrum enhancement   takes place.  It is interesting to better investigate this effect, to understand at what extent the slope of the spectrum has memory of all non-slow-roll eras occurred during inflation.
  
\end{itemize}

 Motivated by these considerations,  
 we develop an analytical, model independent method for studying the spectrum of cosmological fluctuations in single-field inflationary scenarios that include brief, transient departures from slow-roll evolution.    
Calling $\Delta \tau$ the quantity parameterizing the duration of the non-slow-roll epoch, we
organize the evolution equations of fluctuations in a Taylor series in this quantity, and we  
 solve the equations at first order   in $\Delta \tau$.
 The corresponding fluctuation  spectrum   
  is described by few parameters, and presents a pronounced dip followed
  by a rapid growth in amplitude, as  typically found  
   in numerical as well as analytical studies of  non-slow-roll scenarios. The simplicity of our expressions suggests
   new quantitative relations between the features of the  spectrum.  For example, the dip position $k_{\rm dip}/k_*$
    and  the maximal enhancement $\Pi_{\rm max}$ of the spectrum towards small scales are found to be related by
    the  law $k_{\rm dip}/k_*\propto 
    \Pi_{\rm max}^{-1/4}    $, and we determine the proportionality constant. 
     For a single epoch of slow-roll violation 
   we confirm previous studies \cite{Byrnes:2018txb,Carrilho:2019oqg,Ozsoy:2019lyy} finding that  the steepest
   slope of the spectrum has spectral index  $n-1\,=\,4$.   When considering  multiple phases of slow-roll violation,  however, the   slope of the spectrum 
      can be   enhanced.
   This indicates  that
    the  slope of the spectrum has memory  of the  history 
   of non-slow-roll phases during inflation. 
   
   \smallskip
   
   Our work is organized as follows:
   \begin{itemize}
   \item In section \ref{sec-anfor} we develop our analytical  method,   based on a gradient expansion of the perturbation
   equations   and on a Taylor series on the duration $\Delta \tau$ of the non-slow-roll era. After
   introductory sections with our motivations (sections \ref{sec-stcase} and \ref{sec-tranmot}), 
   we 
   analytically solve the
     system  of equations
        at first order in a $\Delta \tau$ expansion, leading to a solution
   of the mode function that depends on  few parameters (sections \ref{sec-ans1}-\ref{sec-resum}). Section
   \ref{sec-compa} briefly compares our approach with other analytical methods discussed in the literature.
   \item In section \ref{sec-PS1} we apply our analytic results to compute the power spectrum of fluctuations and the 
   corresponding spectral index in models of single field inflation with brief phases
   of non-slow-roll evolution. In sections \ref{sec1pssv}, \ref{sec1-match} we focus on scenarios characterized by a single non-slow-roll epoch:
   we  determine new analytic formulas characterizing the properties of  the spectrum, and confirm previous results on slope of the spectral
   index. Section \ref{secMNS} instead considers for the first time scenarios with multiple non-slow-roll epochs, nested between phases of quasi-de Sitter expansion. We find that the slope of the 
    small scale 
   spectrum    towards
   the peak can be  steeper than 
    in the previous case, and we discuss the physical implications of our results for the induced
    spectrum of gravitational waves.
   \item Section \ref{sec-concl} contains our conclusions, followed by  Appendix \ref{App_toymodel} presenting an explicit model
  of non-slow-roll inflation where our formulas can be applied, and by Appendix \ref{App_comparison} discussing and deriving the results
  of section \ref{secMNS} using alternative approaches.
     \end{itemize}

\section{Analytic formulas   for the mode function of fluctuations\\ during   transient epochs of  slow-roll violation}
\label{sec-anfor}

\subsection{The standard case}
\label{sec-stcase}

 We focus on an inflationary epoch of cosmological expansion in single-field inflation, during which for most of the process the metric
 is well approximated by  pure de Sitter space
  \be\label{puredS1}
 d s^2\,=\,a^2(\tau) \left(-d \tau^2+d \vec x^2\right)\hskip0.8cm,\hskip0.8cm a(\tau)\,=\,-\frac{1}{H_0 \tau}\hskip0.8cm,\hskip0.8cm  \tau\le 0\,\,,
 \ee
 up to subleading slow-roll corrections, that we  neglect in this work. Here,
  $\tau$ is the conformal time and $H_0$ the (nearly) constant Hubble parameter during inflation.
 We  consider transient departures from the previous metric, and
we 
 are interested in studying the statistical distribution of the  massless fluctuations of fields
 propagating in this space-time.
     Such fluctuations are controlled
  by the following quadratic action in Fourier space:
\be\label{gen-act}
S\,=\,\frac12\,\int d \tau\, d^3 k\,z^2(\tau)\,\left[ \psi'^2_k(\tau) +k^2\,\psi_k^2(\tau) \right]\,.
\ee
The function $z(\tau)$ is dubbed {\it pump field}: it is model dependent, and characterises the dynamics
of the fluctuations under study.
The structure of the previous quadratic action is  generic in single field inflation. It can describe
the dynamics of the Mukhanov-Sasaki variable $\zeta_k$ controlling curvature  fluctuations
in single-field vanilla models of inflation, where 
$z \propto a\,\dot \phi/H_0$ with $\phi$ the inflaton field; or  the dynamics of massless spin-2 (tensor) or spin-0 (scalar) fluctuations around  pure de Sitter, and then 
$z \propto  a$ (see e.g. \cite{Mukhanov:2005sc}).  But more general possibilities can occur, which are interesting to investigate.
  In a regime of slow-roll single-field inflation, the pump field    has the    generic
profile
\be \label{srpf}
z_{dS}(\tau)\,=\,-\frac{c_0}{H_0\,\tau}\,,
\ee
up to small subleading slow-roll corrections, which  we  neglect in this work. Here, $c_0$ denotes
 a dimensionless quantity  
  depending on the system one considers. For example, when  analysing massless
 fluctuations for a scalar field  in pure de Sitter, one has $c_0\,=\,1$; while for  
 the Mukhanov-Sasaki variable in single-field slow-roll inflation
   $\zeta_k$ one gets $c_0\,=\,\dot \phi/H$. 
  Working with a pump field
as in eq. \eqref{srpf} it is straightforward to solve exactly the evolution equation for
the mode $\psi_k(\tau)$ in this pure de Sitter limit of slow-roll inflation.  We
get
\be\label{phidsG}
\psi_k(\tau)\,=\,-\frac{i}{\sqrt{2 k^3}}\,
\frac{H_0}{c_0}\,
  \left[
{\cal C}_1 \left(
1+i \,k \tau
\right) e^{-i k \tau}
+{\cal C}_2 \left(
1-i \,k \tau
\right) e^{i k \tau}
\right]\,,
\ee
for two arbitrary complex constants ${\cal C}_{1,2}$,
independent from time. 
 The solution that matches the Bunch-Davies vacuum at early times selects ${\cal C}_{1}\,=\,1$, 
  ${\cal C}_{2}\,=\,0$ and 
reads (see e.g. \cite{Mukhanov:2005sc})
\be\label{phids}
\psi_k(\tau)\,=\,-{i}\,\frac{e^{-i k \tau}}{\sqrt{2 k^3}}
\,\frac{H_0}{c_0}\,
 \left(
1+i \,k \tau\right)\,.
\ee

\subsection{Transient deviations from  slow-roll evolution:
\\
our motivations and our approach}
\label{sec-tranmot}

We  
investigate general properties of the spectrum 
   of fluctuations  
  $\psi_k(\tau)$ in scenarios where there exist
 {\it brief, transient epochs} of violation of slow-roll conditions.
    During the brief non-slow-roll epochs the pump field
$z(\tau)$ is {\it not} described by the simple profile of eq. \eqref{srpf}.  In
these phases we do not expect  a regime of  
attractor inflation, hence the would-be decaying mode becomes temporarily important, affecting the 
features of the power spectrum.

 One  motivation for considering this possibility is to amplify the spectrum
 of scalar curvature fluctuations at small scales, so to overcome the threshold of 
 formation for  primordial black holes within single field inflation \cite{Garcia-Bellido:2017mdw,Ezquiaga:2017fvi,Ballesteros:2017fsr,Hertzberg:2017dkh,Motohashi:2017kbs} (see  \cite{Carr:2016drx,Sasaki:2018dmp} for reviews). Another   
   motivation is to investigate models of inflation that amplify
   the amplitude primordial tensor  at frequencies 
  directly testable with gravitational wave experiments (see e.g. \cite{Bartolo:2016ami} for a study focussed
 on LISA).   
 Finally, possible anomalous features in the spectra of CMB temperature fluctuations and spectral distortions could be explained by
   a temporary growth of some of the slow-roll parameters, see e.g.  \cite{Chluba:2015bqa} for a detailed review. 
 
 \smallskip
 
  So far, various models achieve these aims by designing a short phase
  of inflationary non-attractor  evolution. In one class of models,  the first slow-roll
  parameter $\epsilon_1\,=\,-\dot{H}/H^2$ remains small,
  while the absolute value of the second slow-roll parameter $\epsilon_2\,=\,\dot{\epsilon}_1/(H\,\epsilon_1)$
  is large, see e.g. \cite{Kinney:2005vj,Garcia-Bellido:2017mdw,Germani:2017bcs,Motohashi:2017kbs,Dimopoulos:2017ged,Yi:2017mxs,Pattison:2018bct,Cheng:2018qof}.  
   Non-attractor inflation is typically obtained when 
   the pump field entering in eq. \eqref{gen-act} decreases with time, instead of increasing 
  as what happens in the standard case (see eq. \eqref{srpf}, and recall that time runs over negative
  values). 
   A  well studied example of non-attractor evolution is  ultra-slow-roll inflation,
    with $\epsilon_2\,=\,-6$ and $\epsilon_1\sim0$, and pump field $z(\tau) \propto \tau^2$.  Such scenario
     can be well realized in single-field scenarios with inflection points in the inflationary potential (see e.g.
  \cite{Inoue:2001zt,Linde:2001ae,Kinney:2005vj,Martin:2012pe,Motohashi:2014ppa}), where $V_{,\phi}\,\simeq0$, $\epsilon_1$ very small, 
  hence
  \be \label{defsr1}
  \frac{\ddot \phi}{H\,\dot \phi}\,=\,-\epsilon_1+\frac{\epsilon_2}{2}\,\simeq\,-3\hskip0.5cm\Rightarrow
  \hskip0.5cm\epsilon_2\simeq-6\,.
  \ee
    The corresponding
    statistics of fluctuations is related
    to the standard slow-roll case by a duality  \cite{Wands:1998yp} which   leads to   simple, physically
    transparent   formulas for  the spectrum 
    of fluctuations (this approach initially  developed for the scalar sector of curvature fluctuations can  be extended to the tensor sector as well \cite{Mylova:2018yap,Ozsoy:2019slf}).

Another class of models     include 
    scenarios of `punctuated' inflation, where
   brief interruptions of quasi-de Sitter expansion lead to temporary  enhancements  also of the  slow-roll parameter $\epsilon_1$, that then returns to  small values for 
    a final phase of quasi-de Sitter evolution   (see e.g. \cite{Roberts:1994ap,Seto:1999jc,Leach:2000yw,Jain:2007au,Jain:2008dw} for early works and  the review \cite{Chluba:2015bqa}).
     These set-up can be realized
      in systems whose potentials have steps, or discontinuities in their derivatives
    (see e.g. the early papers \cite{Starobinsky:1992ts,Adams:1997de}).
           In this case, 
    analytic approaches are not always available, but numerical methods can be used
    (see e.g \cite{Adams:2001vc}).
  
 \smallskip

 In this work, instead of studying specific models, 
 we enquire whether there are  
    {\it common features} in the properties
    of the power spectrum of fluctuations as a function of the scale, in
       scenarios that include  {\it brief and transient} phases of 
      non-slow-roll evolution 
      \footnote{The condition  of brief 
    non-attractor epoch is  needed for avoiding issues
     with quantum diffusion, see e.g. \cite{Biagetti:2018pjj,Ezquiaga:2018gbw,Firouzjahi:2018vet,Kuhnel:2019xes}.
        }.
     For this aim, we do not use arguments based on slow-roll 
 expansions, nor need to specify explicit values for slow-roll parameters. Instead, we directly work in terms
 of the pump field $z(\tau)$. 
We develop an analytical   method  to solve the mode equations for the perturbation $\psi_k(\tau)$ in the limit of 
    short duration $\Delta \tau$ of non-slow-roll epoch (in comparison with the total duration of inflation),
   that can be applied
    to  general profiles for the pump field,  regardless of the physical origin of the model. 
          In   the limit 
    of small $\Delta \tau$, we   perturbatively Taylor expand the functions involved in this small
    parameter  controlling the duration of non-attractor epoch.

\begin{figure}[h!]
\centering
 \includegraphics[width = 1. \textwidth]{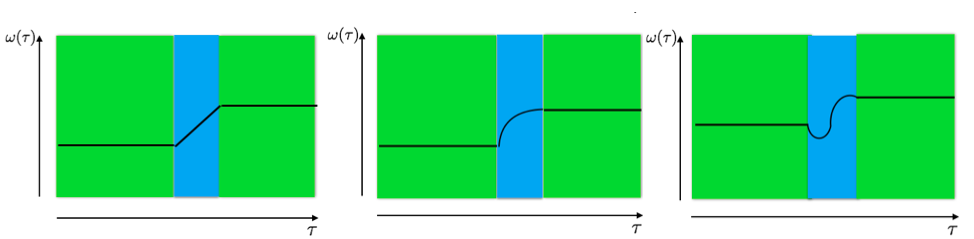}
 \caption{\it  Pictorial representation
  of  arbitrary  profiles for the function $\omega$ in eq \eqref{zgen} which can give rise to brief periods
  of slow-roll violation (in blue in the figure).}
 \label{fig:artistic}
\end{figure}

    We  concretely   develop this plan in what follows. First, we fix some notation. We parameterize 
 the pump field as depending on a function $\omega(\tau)$  (the square
root is included to simplify the formulas we  meet in the next sections)

 \be
 z(\tau)\,=\,\left\{    
 \begin{array}{rl}
  -\sqrt{\omega(\tau_1)}\,\frac{c_0}{H_0\,\tau}
 &\hskip0.6cm
 {\rm for}\hskip0.6cm
 \tau\,<\,\tau_1 
\,,
  \\
 \\ 
 -\sqrt{\omega(\tau)}\,\frac{c_0}{H_0\,\tau}\,
   &\hskip0.6cm
 {\rm for}\hskip0.6cm
 \tau_1\le \tau \le \tau_2\,,
  \\
 \\
   -\sqrt{\omega(\tau_2)}\,\frac{c_0}{H_0\,\tau}
 &\hskip0.6cm
 {\rm for}\hskip0.6cm
  \tau >  \tau_2\,.
  \end{array}\right. \,
  \label{zgen}
 \ee
 
 \noindent
 Without loosing in generality, we  choose $\omega(\tau_1)\,=\,1$. Non--slow-roll evolution
  occurs in the time interval $ \tau_1\le \tau \le \tau_2$. We denote such interval as
    \be \label{intdef}
    \Delta \tau\,\equiv\,\tau_2-\tau_1\,.
    \ee
 The pump field $z(\tau)$ is a continuous function
 of time, but its first derivatives can be discontinuous within the 
    interval $\Delta \tau$.
         In the ultra-slow-roll example discussed above, $z(\tau)\propto \tau^2$ hence $\omega(\tau)\propto \tau^6$,
    while in the case of slow-roll inflation $\omega(\tau)$ is nearly constant. More
    generally, in
     Fig \ref{fig:artistic} we pictorially represent  arbitrary   profiles for the function $\omega(\tau)$ during the 
     non-attractor evolution.

The conformal time $\tau_1$ denotes the conformal time when the inflationary system first leave the slow-roll 
phase, to enter into the non-slow-roll era. We consider cases in which the duration of non-slow-roll evolution
is {\it much shorter} than the typical time-scales of the systems, that we take  of order $|\tau_1|$. This assumption
is quantified by the inequality~\footnote{We prefer to quantify the  short duration of the non-slow-roll
    era in terms of the combination  ${\Delta \tau}/|\tau_1|$, instead of the corresponding number of e-folds of inflation. This 
    since in general beyond slow-roll we do not have information about the scale factor $a(\tau)$, and the e-fold number $\Delta N$, during the time interval $\Delta \tau$.}
\be\label{smdtr}
\sigma\equiv
\frac{\Delta \tau}{|\tau_1|}
 \,\ll\, 1\,,
\ee
which defines a small parameter $\sigma$, useful for our arguments based on a perturbative expansion.

    \smallskip

While in eq \eqref{zgen} we consider  a single phase of non-slow-roll expansion, it is also
possible to consider situations with multiple non-slow-roll epochs, and with intermediate
phases of slow-roll expansion in between. We postpone  these considerations in section \ref{secMNS}. 
 In what comes next we elaborate an appropriate Ansatz for the mode
function $\psi_k$, which  allows us to solve the corresponding evolution equations 
  in the limit of short duration \eqref{smdtr} of non-slow-roll evolution.

\subsection{An Ansatz to describe the non-slow-roll evolution}
\label{sec-ans1}

 In order 
  to study the system, as customary we rescale $$\psi_k(\tau)\,\equiv\,\frac{\varphi_k(\tau)}{z(\tau)}\,.$$
   The  field equation  for $\varphi_k(\tau)$ reads
\be\label{phieq}
\varphi_k''+\left(k^2-\frac{z''}{z}\right)\,\varphi_k\,=\,0\,,
\ee
where we assume a profile for the pump field $z(\tau)$  as   in eq. \eqref{zgen}.\smallskip

It is not immediately clear how to design a method to `separate' the dependence from time $\tau$ and scale
$k$ in eq \eqref{phieq},  and  Taylor expand in small time intervals $\Delta \tau$. We proceed as follows.
  We implement a gradient expansion and adopt the following perturbative  Ansatz
\bea\label{phians}
\varphi_k(\tau)&=&-i\,\frac{e^{-i k \tau}}{\sqrt{2 k^3}}\,{ z(\tau)}\,
\frac{H_0}{c_0}\,
 \,{\cal C}_1\,
\left[
1+i \,k \tau+\left(i k \tau_0\right)^2 \,A_{(2)}(\tau)+\left(i k \tau_0\right)^3\,A_{(3)}(\tau)+\dots
\right]
\nonumber
\\
&&
-i\,\frac{e^{i k \tau}}{\sqrt{2 k^3}}\,{ z(\tau)}
\frac{H_0}{c_0}\,
 \,{\cal C}_2\,
\left[
1-i \,k \tau+\left(-i k \tau_0\right)^2 \,A_{(2)}(\tau)+\left(-i k \tau_0\right)^3\,A_{(3)}(\tau)+\dots
\right]\,,
\nonumber
\\
\eea
which has the property to match to the standard de Sitter mode function when the functions $A_{(n)}=0$. 
Recall that the overall factor $H_0/c_0$ (introduced in eq \eqref{srpf}) accommodates   distinct possibilities for the nature of fluctuation $\psi_k$ (scalar fluctuation, Sasaki-Mukhanov variable, tensor modes etc). 
 The previous Ansatz contains 
 a new time scale $\tau_0$ that we   use for  
defining a  convenient  dimensionless parameter $\Delta \tau/\tau_0$. The final results are independent
from $\tau_0$ though. 

In what comes next,
the dimensionless functions $A_{(n)}$ entering eq \eqref{phians} are   chosen   so  to ensure  
that they
vanish 
for $\tau\le\tau_1$. This implies that  we 
 get the general de Sitter solution of eq \eqref{phidsG} at early times: if we wish to match with the Bunch-Davies vacuum, then
${\cal C}_1=1$ and ${\cal C}_2=0$; but for our discussion we  keep arbitrary values for these constants. 
 The quantity $\tau_0$ is the reference scale \eqref{srpf}. Similar  perturbative approaches
  have been developed in other works, as we  briefly  review in section \ref{sec-compa}.

\smallskip

Plugging the   Ansatz \eqref{phians} into eq \eqref{phieq}, we  analyze  the problem  in terms of a gradient expansion in powers of $k$. We find 
the following system of differential equations in the time coordinate, valid for each power of $k$ bigger or equal to 2 (a prime indicates derivative along time):
\bea
\label{eqA2}
\left[ \frac{\omega(\tau)}{\tau^2}\,\tau_0^2\,A_{(2)}'(\tau)
\right]'&=&\frac{\omega'(\tau)}{\tau}\,,
\\
\left[ \frac{\omega(\tau)}{\tau^2}\,
\left(
\tau_0\,A_{(n)}'(\tau)-A_{(n-1)}(\tau)
\right)
\right]'&=&\frac{\omega(\tau)\,A_{(n-1)}'(\tau)}{\tau^2}
\hskip0.3cm, \hskip0.3cm{\text{\,for $n>2$}}\,.
\label{eqAn}
\eea

\noindent 
When $\omega(\tau)$ is a constant we learn that  $A_{(n)}=0$ is a consistent solution 
of the previous system of equations; on the other
hand,  any non-trivial profile of $\omega(\tau)$ 
switches on the $A_{(n)}$-contributions.

In fact,
a formal solution for $A_{(2)}$ in eq \eqref{eqA2} is given
by the expression
\be\label{solA2t}
\tau_0^2\,A_{(2)}(\tau)\,=\,\int_{-\infty}^{\tau}\,d \tau_a\,\frac{\tau_a^2}{\omega(\tau_a)}\,
\left(
\int_{-\infty}^{\tau_a}\,d \tau_b\,\frac{\omega'(\tau_b)}{\tau_b}
\right)\,.
\ee
 The extremes of integration   are chosen
 to satisfy the conditions we wish to impose: when $\tau\le\tau_1$, then   $A_{(2)}(\tau)=0$.
 Alternatively, we can set the lower limits of the integral to $\tau_1$, and set  $A_{(2)}(\tau)=0$
 for $\tau\le\tau_1$. This choice of  boundaries completely fix  the solution for $A_{(2)}(\tau)$.

Passing to higher $n$'s, 
a formal solution for $A_{(n)}$ in eq \eqref{eqAn} is given
by the expression
\be
\tau_0\,A_{(n)}(\tau)\,=\,\int^{\tau}_{-\infty}\,d \tau_a\,A_{(n-1)}(\tau_a)
+\int_{-\infty}^{\tau}\,d \tau_a\,\frac{\tau_a^2}{\omega(\tau_a)}\,
\left(
\int_{-\infty}^{\tau_a}\,d \tau_b\,\frac{\omega(\tau_b)\,A_{(n-1)}'(\tau_b)}{\tau_b^2}
\right)\label{forsoln}\,.
\ee
Again, the extremes of integration are  chosen  to ensure that
 $A_{(n)}(\tau)\,=\,0$ for $\tau\le \tau_1$.

\smallskip

This is the first new result of our work:
 we determined the {\it most general} formal solutions for the functions $A_{(n)}$. These formal solutions
satisfy the correct boundary conditions that completely specify the system. If we were able to systematically
 perform all the integrals in eqs \eqref{solA2t}, \eqref{forsoln}, 
we would get the full solution for the equation \eqref{phieq}  we wish to solve.
  We
 now exploit the fact that the  epoch of slow-roll violation lasts a short time interval $\Delta \tau\,=\,\tau_2-\tau_1$
   and we compute the exact expressions for the $A_{(n)}$  ($n\ge2$) at leading order in $\Delta \tau/\tau_0$.

\subsection{Solving the equations at leading order in $\Delta \tau/\tau_0$}
\label{sec-solv2}

 Working at leading
 order in a Taylor expansion in $\Delta \tau/\tau_0$, 
 we can explicitly  determine   the solutions for the functions $A_{(n)}$ (with $n\ge2$).
   Since  $\Delta \tau$ is small, we can 
  Taylor expand, 
  and keep the first non-vanishing term. 
 Hence, for the function $A_{(2)}$ we find
\be
A_{(2)}(\tau_2)\,=\,A_{(2)}(\tau_1)+A^{\bf 1}_{(2)}(\tau_1)\Delta \tau+\frac12 A^{\bf 2}_{(2)}(\tau_1) \Delta \tau^2+\dots
\ee
where the {\it upper index in boldface} indicates from now on the order  of derivative. It is clear that $A_{(2)}(\tau_1)=0$. Then using~\footnote{These arguments are valid  in the
physically reasonable assumption that the derivatives of $\omega(\tau)$ do not 
diverge within the small interval $\Delta \tau$, so that $\lim_{\Delta \tau\to0} \omega^{\bf n}(\tau_1) \,\Delta \tau\,=\,0$.
\label{footbound}}
  formula \eqref{solA2t}
   we get
\be
\tau_0^2\,A^{\bf 1}_{(2)}(\tau)\,=\,\frac{\tau^2}{\omega(\tau)}\,
\int_{-\infty}^{\tau}\,d \tau_a\,\frac{\omega'(\tau_a)}{\tau_a}\hskip1cm\Rightarrow \hskip1cm A^{\bf 1}_{(2)}(\tau_1)\,=\,0\,,
\ee
due to the fact that for $\tau<\tau_1$ we have $\omega'=0$. 
On the other hand,   the second derivative has  a non-vanishing contribution:
\be\label{aldef}
\tau_0^2\,
A^{\bf 2}_{(2)}(\tau_1)\,\equiv\,\alpha\,=\,\frac{d\,\ln \omega(\tau)}{d \ln \tau}\,\Big|_{\tau=\tau_1}\,.
\ee
This implies that the leading contribution to the Taylor series  in an expansion in $ \frac{\Delta \tau}{\tau_0}$
is
\be\label{leadA2}
A_{(2)}(\tau_2)\,\simeq\,\frac{\alpha}{2}   \frac{\Delta \tau^2}{\tau_0^2}\,.
\ee
 We can proceed in the same spirit with $A_{(3)}$. We have
\be
\tau_0\,A_{(3)}^{\bf 1}(\tau)\,=\,A_{(2)}(\tau)+
 \frac{\tau^2}{\omega(\tau)}\,
\int_{-\infty}^{\tau}\,d \tau_a\,\frac{\omega(\tau_a)\,A_{(2)}^{\bf 1}(\tau_a)}{\tau_a^2}\,.
\ee
Simple manipulations, using also the previous results \eqref{aldef},  \eqref{leadA2} obtained for $A_{(2)}(\tau_1)$ and 
its derivatives, lead to 
\bea
A_{(3)}^{\bf 0}(\tau_1)&=&A_{(3)}^{\bf 1}(\tau_1)\,=\,A_{(3)}^{\bf 2}(\tau_1)\,=\,0\,,
\\
\tau_0^3\,A_{(3)}^{\bf 3}(\tau_1)&=&2\,\alpha\,.
\eea
Proceeding recursively, one finds that 
 the first  
 non-zero derivative 
 of $A_{(n)}$ evaluated 
 at $\tau=\tau_1$ is the $n$-th, with
\bea
\tau_0^n \,A_{(n)}^{\bf n}(\tau_1)&=&2\,\tau_0^{n-1} \,A_{(n-1)}^{\bf {n-1}}(\tau_1)\,=\,2^{n-2} \,\alpha\,.
\eea
This implies that for each $n$, the leading order in the corresponding Taylor expansion 
 is 
\be\label{solAN1}
A_{(n)}(\tau_2)\,\simeq\,\frac{2^{n-2}}{n!} \,\alpha\, \frac{\Delta \tau^n}{\tau_0^n}\,, \hskip1cm
{\text{for any $n\ge2$. }}
\ee
 Hence,
 we have been able to solve at each order in $n$ the system of equations
\eqref{eqA2}, \eqref{eqAn}, determining for each $n$ the leading order contribution to  the functions
$A_{(n)}$, in a Taylor expansion in the  quantity $\Delta \tau/\tau_0$. The solutions depend on the
parameter $\alpha$ -- defined in  \eqref{aldef} -- characterising the effective slope of the pump field $z(\tau)$, as
well as on the duration $\Delta \tau$
of the non-slow-roll era.

\smallskip

How much are we loosing neglecting the higher order terms in each of  the Taylor expansions discussed
above? Consider for example $A_2(\tau)$. Including the next-to-leading contribution to
the Taylor expansion we find a correction to eq. \eqref{leadA2} which reads 
\be\label{A2t2c}
A_{(2)}(\tau_2)\,=\,\frac{\alpha}{2}   \frac{\Delta \tau^2}{\tau_0^2}\left[
1+\frac{\Delta \tau}{3 \,\tau_1} \left( 
2-\alpha+\frac{d\,\ln \alpha}{d\,\ln \tau}
\right)
 \right]+\dots
\ee
where all quantities in the right-hand-side of the previous expressions are evaluated at $\tau_1$.  More
generally, all the remaining corrections are weighted by powers of $\Delta \tau/\tau_1$. We  impose
the hierarchy $|\Delta \tau/\tau_1|\ll1$: 
 this corresponds to take
a small parameter $\sigma$ in
 eq. \eqref{smdtr}, and this 
 choice is motivated by the short  duration of the
non-slow-roll epoch with respect  with the   time-scale $|\tau_1|$. (Instead, the time-parameter 
  $\tau_0$ can be selected  so that the ratio  $\Delta \tau/\tau_0$ is  
parametrically larger than $\Delta \tau/\tau_1$, so that the leading term in the expansion is non-negligible.) We also assume that  the small quantity
$\Delta \tau/\tau_1$ in eq. \eqref{A2t2c} is not much amplified  by  enhancing factors, as the
ones contained in the round brackets  of eq.  \eqref{A2t2c} in the limit of large parameter $\alpha$. 
 This procedure defines
a perturbative scheme that includes only the leading contributions weighted
by powers of $(\Delta \tau/\tau_0)^n$ in  \eqref{solAN1},  while it 
neglects all the other terms in the Taylor series.

\subsection{Resumming the series, and the leading expression for the mode function}
\label{sec-resum}

We can plug these results in the solution for the mode function $\psi_k\,=\,\varphi_k/z(\tau)$,
 evaluating it at any time  $\tau$ within the interval $ \tau_1\le \tau \le \tau_2$ 
when  a violation of slow-roll conditions occurs. As an example, we focus on  $\tau\to\tau_2$
i.e. towards the end of the non-slow-roll era.
 Setting ${\cal C}_1=1$ and ${\cal C}_2=0$ in  
eq \eqref{phians}, so to select a Bunch-Davies vacuum at early times, we get 
\bea
\psi_k(\tau_2)&\sim&-i\,\frac{e^{-i k \tau_2}}{\sqrt{2 k^3}}\,\frac{H_0}{c_0}\,\left[
1+i \,k \tau_2+\left(i k \tau_0\right)^2 \,A_{(2)}(\tau_2)+\left(i k \tau_0\right)^3\,A_{(3)}(\tau_2)+\dots
\right]\,,
\nonumber
\\
&=&-i\,\frac{e^{-i k \tau_2}}{\sqrt{2 k^3}}\,\frac{H_0}{c_0}\,
\left[
1+i \,k \tau_2-\frac{\alpha }{4}-\frac12\,i\,k\,\alpha\, \Delta \tau+\frac{\alpha }{4}\,\sum_{n=0}^{\infty}\,\frac{(2\,i\, k)^n}{n!}\,\Delta \tau^n
\right]\,.
\eea
We can resum exactly the series, and get 
   (recall that $\Delta \tau = \tau_2-\tau_1$)
\bea
\psi_k(\tau_2)
&=&-i\,\frac{e^{-i k \tau_2}}{\sqrt{2 k^3}}
\,\frac{H_0}{c_0}\,
 \left[ \left(
1+i \,k \tau_2 \right)
 -\frac{\alpha }{4} \left( 1+2\,i\,k\, \Delta \tau
-\,e^{2 i k \Delta \tau} \right)
\right]\label{analytic1}\,,
\eea
for the mode function at the end of the non-slow-roll epoch.  

The previous formula can be easily generalised
to any value of ${\cal C}_{1,2}$ -- hence allowing for a richer mode evolution or a more general vacuum preceding  the non-slow-roll epoch -- leading to 
\bea
\psi_k(\tau_2)
&=&-i\,\frac{e^{-i k \tau_2}}{\sqrt{2 k^3}}\,{{\cal C}_1}
 \,\frac{H_0}{c_0}\,
\left[ \left(
1+i \,k \tau_2 \right)
 -\frac{\alpha }{4} \left( 1+2\,i\,k\, \Delta \tau
-\,e^{2 i k \Delta \tau} \right)
\right]
\nonumber
\\
&&-i\,\frac{e^{i k \tau_2}}{\sqrt{2 k^3}}\,{{\cal C}_2}
 \,\frac{H_0}{c_0}\,
\left[ \left(
1-i \,k \tau_2 \right)
 -\frac{\alpha }{4} \left( 1-2\,i\,k\, \Delta \tau
-\,e^{-2 i k \Delta \tau} \right)
\right]\,.
\label{analytic1a}
\eea

These relatively simple formulas are  at the basis of our analytical considerations.  
  The mode function \eqref{analytic1} (or \eqref{analytic1a})
 {\it can not} be recasted in terms of the general solution    \eqref{phidsG} for massless modes in de Sitter space
 with a simple profile \eqref{srpf} for the pump function. 
 The correction to the `pure de Sitter' solution of eq \eqref{phids} is weighted by
 the dimensionless parameter $\alpha$
 within the square parenthesis.
   The results so far have their points of interest and their limitations:
 \begin{itemize}
 \item The mode function \eqref{analytic1} modifying  the de Sitter solution is extremely simple. It depends  on the  effective parameter
 $\alpha$, which condenses in a single quantity any deviation 
   with respect to the de Sitter case, and on the duration
$\Delta \tau$ of the   non-slow-roll epoch. 
 It  can be applied to parameterize different scenarios with brief violations of slow-roll conditions. For the case of ultra-slow-roll expansion, 
 one has $\omega(\tau) \propto \tau^6$  and $\alpha\,=\,6$, but  also other   values  are allowed during non-slow-roll inflation, depending on the scenarios~\footnote{
 Appendix \ref{App_toymodel} discusses
 a toy model of single-field inflation with an arbitrarily fast decreasing pump field,
   by suitably tuning the inflationary potential. This
 shows that, at least in principle,  any value of $\alpha$ can be achieved in single-field inflation.}.
  In our study,  $\alpha$ is  considered a free  parameter, that can assume large values. 
  The possibility to study the system  in terms of  very few effective parameters has its clear advantages.
   For example,   in what next we  show that we can combine our results and get universal formulas connecting properties
   of the power spectrum as a function of the scale, which are independent from   specific  models. 
\item On the other hand,
it is important to stress that we are {\it truncating} each of the Taylor series at the first non-vanishing contribution, and we are not always ensured that the following terms are smaller than the first in all models under consideration.  In the previous section \ref{sec-solv2} we identified a perturbative procedure that ensures  that  our truncation leads to reliable
results, if the non-slow-roll period $\Delta \tau$ is short, and the system satisfies appropriate conditions. (See in
particular the discussion after eq. \eqref{A2t2c}.) 
 On the other hand, models can exist where $\Delta \tau$ is not negligibly small:
 the higher order, next-to-leading Taylor corrections to our results
  can be relevant,  if 
  they
   are weighted by  large coefficients that amplifies them. 
    In our investigation we do not enter into explicit analysis
   of specific models, and we assume to work within systems where our perturbative approach is valid, and our
   approximations correct. In general, we expect that
  our perturbative scheme   works best  when the growth
 of the spectrum happens  quickly  through   sudden variations of the slow-roll parameters.

  \end{itemize}

\subsection{Brief comparison with previous analytical analysis of the spectrum }
\label{sec-compa}

Before
    starting to analytically explore the consequences
     of 
     eq.  \eqref{analytic1}, we conclude this Section
 comparing 
   our procedure with  other analytic approaches proposed  in
       the literature to study the dynamics of
        fluctuations in  inflationary scenarios
      containing phases of non-slow-roll  evolution. 
      
      The work     \cite{Byrnes:2018txb} uses an analytical 
              method  based on junction conditions,  and requires the   knowledge of mode functions during  the non-attractor era. 
   \cite{Byrnes:2018txb} models the non-attractor evolution
      in terms of a combination of  phases,   
each with  %
   constant   slow-roll parameter $\epsilon_2$ (that can possibly be large, up to order $|\epsilon_2|\sim {\cal O}(10)$)
 and small slow-roll parameter $\epsilon_1$. For example, choosing $\epsilon_2\,=\,-6$
 corresponds to   
  ultra-slow-roll  (USR) inflation.
     The mode functions computed in USR and de Sitter phases as
  $\psi_k^{\rm USR}(\tau)\,\propto\,a^{3}(\tau) \,\psi_k^{\rm dS}(\tau)$, hence  during USR the amplitude of the  spectrum of fluctuations 
  grows with time as the sixth power of the scale factor. (Notice that this is  different from our mode
  function \eqref{analytic1}).
     The total duration of non-slow-roll
  epochs last few e-folds, in order to obtain the necessary amplification of fluctuations.  
   The
 paper \cite{Carrilho:2019oqg} works in the  same context and under the same hypothesis of 
 \cite{Byrnes:2018txb}, providing an interpretation for the steepest growth of the spectrum in such scenarios, and
   additionally showing that a prolonged phase of non-slow-roll expansion preceding 
       a phase of ultra-slow-roll inflation can enhance the slope of the spectrum of perturbations at superhorizon
       scales.

\smallskip

An alternative, systematic approach more similar to ours was proposed in \cite{Leach:2001zf}, 
solving the mode functions
 implementing a gradient  
 expansion in powers of $k$ (although using a slightly different Ansatz with respect to us)
and
studying the role of the decaying mode after horizon crossing. 
The motivation of this work is to study set-up where more drastic departures from slow-roll inflation
occur, with the
   main purpose  to find the conditions for obtaining a growth of the spectrum of perturbations.
      Their methods was preceded by an early paper
  \cite{Seto:1999jc} that proposed  to use a $k$-expansion in solving the mode equations  in certain
   scenarios of 
  interrupted inflation.
  More recently, the methods of \cite{Leach:2001zf} were reconsidered in \cite{Ozsoy:2019lyy} by 
   including 
  higher powers of $k$ in the gradient expansion, so to better understand different slopes of the spectrum 
   of perturbations
  in the process of its
   growth towards its peak. 
  None of these works implemented a truncated Taylor expansion in their analysis, as we do here,
  with the specific purpose to examine what are general features of systems characterised
  by a short duration 
  of the non-slow-roll epoch.

 \section{The power spectrum and its properties}
 \label{sec-PS1}

 In this section we  consider systems where the power spectrum of fluctuations  parametrically
  grows by a factor $\sim 10^5$ from large
towards small scales. This is a representative large quantity aimed at demonstrating that our formalism can be in principle applied to a variety of cases where the amplitude  of perturbations is enhanced during inflation
 -- in the scalar and/or in the tensor sectors \footnote{For the production of primordial
black holes, the  required enhancement of the curvature perturbation spectrum
 from large to small scales is  larger, typically  of order $10^{7}$. Such numbers
  can be obtained in our
 approach by tuning the available 
 parameters.
 }. 
 We are interested in investigating  the corresponding properties of the 
spectrum as a function of the scale.

 We define the power spectrum associated with the mode function $\psi_k(\tau)$ 
 of \eqref{analytic1} at time $\tau$
 as
 \be
 {\cal P}_{k} (\tau)\,=\,\frac{k^3}{2\,\pi^2}\,\langle \psi_k^*(\tau)\,\psi_k(\tau)\rangle'\,,
 \ee 
 where the prime indicates a 2-point function understanding the $\delta-$function associated
 with momentum conservation. Making use of the formulas discussed in the previous section, we 
 can  ensure that all mode functions match a Bunch-Davies vacuum at early times.

    It is  convenient to work with the quantity 
 \be\label{defPI}
 \Pi_{k}(\tau)\,\equiv\,\frac{{\cal P}_{k}(\tau)}{{\cal P}_{k\to0}(\tau)}\,,
 \ee
 the 
 ratio between spectra evaluated respectively at scale $k$ and at very large scales $k\to0$: this ratio goes
 to $1$ for $k\to0$, and 
 makes more manifest the small-scale growth of the spectrum in scenarios with transient violation of slow-roll
 conditions. The  spectral index  -- which generally depends on the scale -- is defined as
 \be
 n-1\,=\,\frac{d \ln \Pi_k(\tau)}{ d \ln k}\,.
 \ee
 An examination of various examples of power spectra profiles for single field inflation discussed in the literature  shows two salient features:
 \begin{enumerate}
 \item While the profile  of $ \Pi_{k}(\tau)$  as a function of $k$ initially gently decreases  as $k$ increase, it typically 
  features a  pronounced dip at some scale $k_{\rm dip}$, associated with modes
  that leave the horizon still during the slow-roll era. This fact
  has been noticed in several papers that numerically analyse the curvature power spectrum in these scenarios, see e.g.  \cite{Garcia-Bellido:2017mdw,Ballesteros:2017fsr,Ezquiaga:2017fvi,Hertzberg:2017dkh,Cicoli:2018asa,Ozsoy:2018flq}   \footnote{More than one dip can be present
  at relatively large scales (see e.g. \cite{Ozsoy:2018flq}) but in the examples we know there is always one dip much more
  pronounced than the others.}. It would be interesting to obtain quantitative,
   model-independent estimates of the position and depth of the dip, and its relation with other features of the spectra. Information on the dip  position  
  allows one to 
  estimate at what scale the spectrum starts to strongly deviate from the amplitude and scale-dependence predicted in single-field slow-roll inflation. This  can be important when comparing models with the constraints on the large
  scale spectra obtained from CMB
  temperature fluctuations and spectral distortions (see e.g. \cite{Bugaev:2010bb,Chluba:2012we,Nakama:2017xvq,Byrnes:2018txb,Sato-Polito:2019hws,Unal:2020mts}). Also, the physics of the dip
  might be associated to a   phase transition, see \cite{Itzhaki:2008hs,Downes:2012xb}.
  \item Soon after the dip, the spectrum starts to steadily increase as $k$ increases, with a slope that normally does not 
   overcome the limit $ n-1\,=\,4$ in single field inflation \cite{Byrnes:2018txb}.  This limit  has nevertheless been shown to be slightly  violated 
  if  a phase of non-attractor
  is immediately preceded  by another phase of slow-roll violation \cite{Carrilho:2019oqg} (see also \cite{Ozsoy:2019lyy}). 
  These facts indicate that the behaviour of the spectrum during phases of non-slow-roll evolution
  might have a `memory' of what occurs prior to these phases, and such memory can be  stored in the slope of the
  spectrum. This information is important for studying possible bounds on the   process of 
  in the amplitude of perturbations.
   For example, a
  stochastic gravitational background induced at second order by enhanced curvature perturbations \cite{Mollerach:2003nq,Ananda:2006af,Baumann:2007zm,Saito:2008jc} can
  be constrained today by 
   pulsar timing array observations\cite{Lentati:2015qwp,Arzoumanian:2020vkk}.
       \end{enumerate}
In what follows we study the previous two points using our formalism, considering two situations. In the first, we assume there is a unique 
phase of brief non-slow-roll evolution, following and preceding two epochs of (quasi)-de Sitter
expansion,  and we seek for  analytic formulas for the properties of the dip and
the slope of the corresponding spectrum of fluctuations.
In the second, we  study scenarios with multiple transient phases of non-slow-roll
evolution, connected
by epochs of slow-roll expansion: we show that the slope of the resulting spectrum can in principle 
increase with respect to scenarios with a single short phase of non-slow-roll evolution.

 \subsection{A single phase of slow-roll violation}
 \subsubsection{The power spectrum during the epoch  of slow-roll violation}
\label{sec1pssv}

 We start studying the case of a single phase of non-slow-roll inflationary evolution, focussing on the spectrum evaluated at a time $\tau$ between times $\tau_1$ and $\tau_2$ (for $\tau\le\tau_1$ and $\tau_2\le\tau\le0$ we assume standard de Sitter expansion). 
   Evaluating the function $\Pi_{k}(\tau)$ of
 eq \eqref{defPI} during such interval
  and denoting $$\Delta \tau_A \,=\,\tau_2-\tau_1\,$$ we get
 
\bea
\Pi_k(\tau)&=&
 1+\tau^2\,{k^2}+
\nonumber
\\
&+&\frac{\alpha}{4}\left[
\left(k\,\Delta \tau_A\,
\right)^2
\left( {\alpha}-\frac{4\,\tau}{\,\Delta \tau_A} \right)-\,
k\,\Delta \tau_A
\,\left(\alpha-\frac{2\,\tau}{\,\Delta \tau_A}  \right)\,\sin{\left({2\,k} \Delta \tau_A\right)}
+\left(\alpha-{4}\right) \,\sin^2{\left({k} \,\Delta \tau_A\right)}
\right]\,.
\nonumber
\\
\label{defPI}
\eea
The contributions due to non-slow-roll epochs are contained in the second line: they vanish at large scales $k\to0$,
and become  important at small scales. 
 The function \eqref{defPI} is an oscillating function with scale-dependent coefficients, which grows unbounded with momentum $k$. This is not surprising since the function
is evaluated at time $\tau$ during the non-slow-roll epoch, and we  do not consider yet a second phase of de Sitter expansion for $\tau\ge \tau_2$ (we  do so in section \ref{sec1-match}).

\begin{figure}[h!]
\centering
 \includegraphics[width = 0.6 \textwidth]{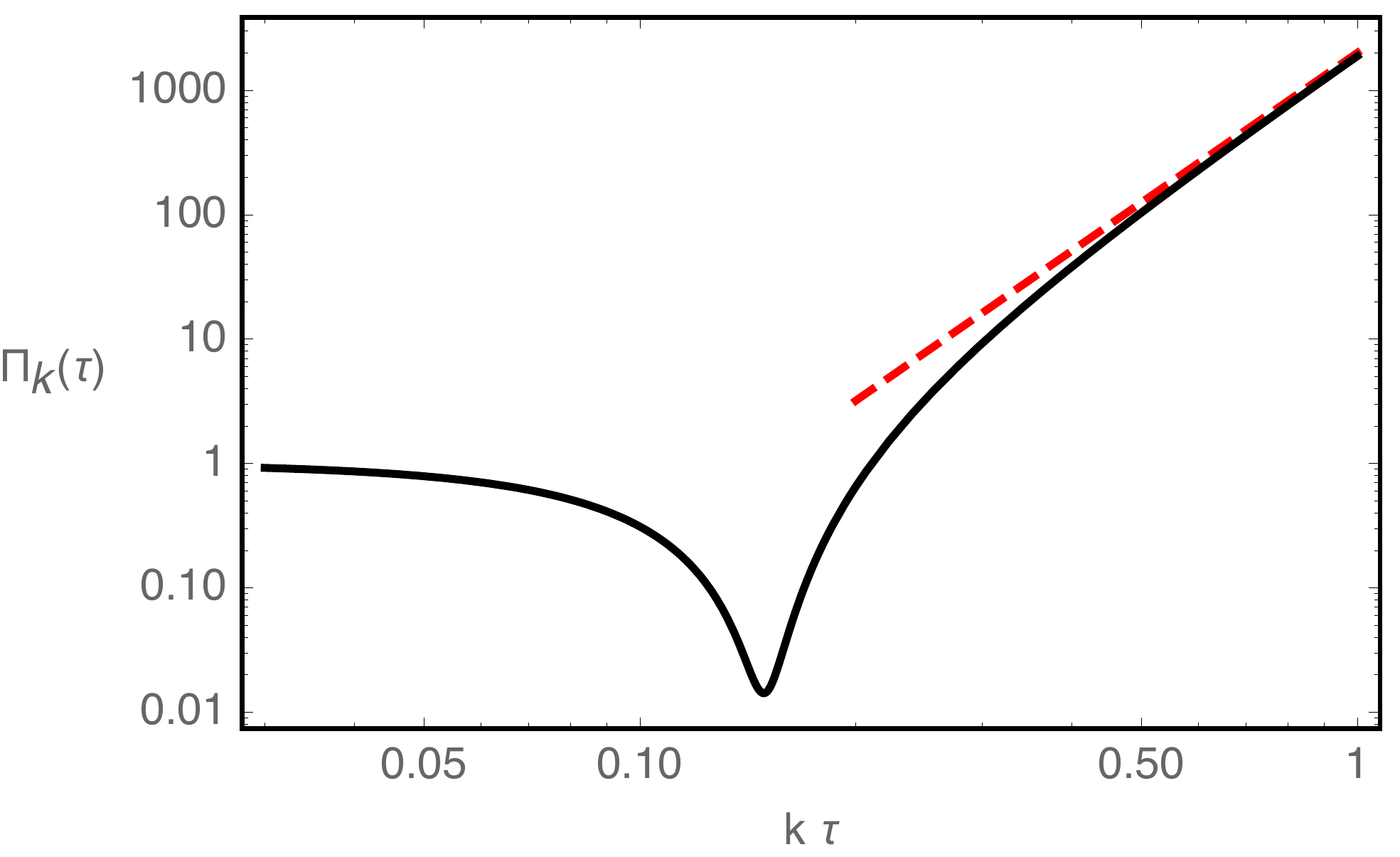}
 \caption{\it Plot of the analytic 
 function $\Pi(k)$ of eq.  \eqref{defPI},  versus the  dimensionless combination $k\, \tau$. 
    We choose
 the representative values $\alpha=1000$, $\Delta \tau_A/\tau\,=\,0.3$ 
 for the parameters that enter 
  in   eq.  \eqref{defPI}. The red dashed line has a profile proportional to  $(k \tau)^4$ and   indicates the maximal
  slope  the spectrum can acquire after the dip.}
 \label{fig:complete-profile-cas1}
\end{figure}

On the other hand, it is instructive to start  by   studying the features
of the spectrum \eqref{defPI}, 
  to point out general properties that are  in common with all the more complex cases we  study in what follows. 

\subsubsection*{The  growth of the power spectrum}

Since there is only a single parameter $\alpha$
which controls any modifications of standard slow-roll formulas, we expect that this parameter controls
the growth of the spectrum: the larger $\alpha$ is, the larger the growth.  In fact, we find  the following expression for the spectrum $\Pi(k)$,
evaluated at a scale $k \tau\sim 1$, in the limit
of large $\alpha$ and small period $\Delta \tau_A$: 
\be\label{pimax}
\Pi_{k=1/\tau}(\tau)\,\equiv\,\Pi_{\rm max}\,\simeq\,\frac{\alpha^2}{4} \left( \frac{\Delta \tau_A}{\tau} \right)^4+\dots
\ee
Of course, evaluating the spectrum at the scale $k \tau\sim 1$ is quite arbitrary at this stage, since
the spectrum grows with no bounds. In the
next sections \ref{sec1-match}
 and \ref{secMNS}, after matching with  phases of quasi-de Sitter expansion, the expressions for
 $\Pi_{\rm max}$ will become  physically more meaningful. 
If we wish to  to gain several orders of magnitude in the amplitude of the spectrum with respect to its large-scale value at $k\to0$, we need to choose  a large  phenomenological parameter $\alpha$.

\subsubsection*{The position and depth of the dip}

Figure \ref{fig:complete-profile-cas1} shows a dip in the spectrum before a rapid growth: this feature is in common with all the scenarios we study
and is  known in numerical \cite{Garcia-Bellido:2017mdw,Ballesteros:2017fsr,Ezquiaga:2017fvi,Hertzberg:2017dkh,Cicoli:2018asa,Ozsoy:2018flq} as well as analytical studies \cite{Leach:2001zf,Byrnes:2018txb}.
  Working in the limit of large parameter $\alpha$, we  analytically understand and analyze
 the dip position and its properties. 
  We expand function \eqref{defPI} at small scales:
   \be\label{smallPk}
\Pi_k (\tau)\,=\,1- \tau^2 k^2  \left(
\alpha\,\frac{ \Delta \tau_A^2}{\tau^2} -1\right)
+{\cal O}\left(  \tau^4 k^4 \right)\,.
\ee
While for $k\to0$ this expression goes to one, for  slightly larger values of $k$ we expect the second term to become important, and (if $\alpha$ is positive) to reduce the amplitude of the spectrum.
 At larger values of $k$ the additional powers $ (k\,\tau)^4$  take over,  leading to a growth in $k$, and causing
  the dip
 we  notice in Fig \ref{fig:complete-profile-cas1}. In fact, we can do better and analytically determine 
the dip position, since the 
 dip is expected to be located at relatively large scales. We rescale the momentum $k$ by an inverse power
 of $\alpha$ introducing the variable $x$ as
 \be\label{anstA} 
\Delta \tau_A\,k\,\equiv\,\frac{\sqrt{2}\,x}{\sqrt{\alpha}}\,.
\ee
 This Ansatz is motivated by the fact that the dip position occurs at relatively large scales, and 
 it is  expected to be controlled by some inverse powers of $\alpha$ -- being this the only 
 potentially large parameter available.  In fact,
we can 
express  the spectrum \eqref{defPI} as a function of  the quantity $x$ defined in eq \eqref{anstA}. By  making an expansion for large $\alpha$ 
  we notice 
that the resulting formula has a leading term, followed by subleading terms that are weighted by inverse powers of 
the large parameter $\alpha$:
\be
\label{PiXexp}
\Pi_x(\tau)\,=\, (x^2-1)^2+\frac{x^2}{9\,\alpha}\frac{18 \tau^2+12 x^2 \Delta \tau_A (\Delta \tau_A-\tau)
 -4\,x^4 \Delta \tau_A^2
 }{\Delta \tau_A}+{\cal O}\left(\frac{1}{\alpha^2}\right)\,.
\ee
 Hence the value $x\,=\,1$  corresponds to  the  position where spectrum exhibits a dip,  being suppressed
by the large value of $\alpha$.  Since $\alpha$ also controls the total growth of the spectrum, we expect that position and amplitude of the dip can be analytically related with some powers of this quantity. 
In fact,  formulas 
\eqref{PiXexp} and \eqref{pimax}  combined together suggest the relations
\be
\Delta \tau_A\,k_{\rm dip}\,\propto\,\left(\frac{1}{\Pi_{\rm max}}\right)^{1/4}\hskip1cm,\hskip1cm\Pi_{k\,=\,k_{\rm dip}}
(\tau)\,\propto\,
\left(\frac{1}{\Pi_{\rm max}}\right)^{1/2}\,,
\ee
that will be made more quantitative in the next sections,  once we include an additional phase of de Sitter expansion.   
 We find these formulas potentially interesting since they are  independent from $\alpha$: they provide universal, model independent relations between the property of the dip and the maximal growth of the spectrum. 

\subsubsection*{The maximal slope of the spectrum in the final phase of growth}

After the dip, the small-scale
 spectrum of Fig  \ref{fig:complete-profile-cas1} 
 starts to steadily grow, with  a spectral index whose 
 maximal value 
   is $n-1=4$, 
     in
agreement with the analysis of \cite{Byrnes:2018txb}.

Notice that the maximal power of $k$
in the coefficients of the oscillating functions of  \eqref{defPI} is $k^2$: on the other hand, the oscillating functions
have their own scale-dependence that can `reinforce' the slope of the spectrum, allowing for higher
powers of $k$.   
  We can understand analytically these facts from our analytic formula \eqref{defPI}. We
  focus at small scales  far away from the dip, and far  from the large scale regime where a large $\alpha$ expansion can 
  be compensated by very small values of $k \,\tau$.  
 Working at such  small  scales, we   concentrate  on the leading order contributions to the spectral index in an expansion in inverse powers of $1/\alpha$. 
  We find, expanding for large $\alpha$,
\bea
n-1&=&
\frac{8 \,( k \,\Delta \tau_A)^2\,\sin^{2}( k \,\Delta \tau_A) }{1+2 \,( k \,\Delta \tau_A)^2-\cos{(2\, k \,\Delta \tau_A)} -2 \, k \,\Delta \tau_A\, \sin{(2 \, k \,\Delta \tau_A)}} +{\cal O}\left(\frac{1}{\alpha}\right)
\nonumber
\,,
\\
&=&4-\frac{4}{9}\,( k \,\Delta \tau_A)^2+{\cal O}\left(\frac{1}{\alpha}, \,( k \,\Delta \tau_A)^4\right)\,.
\label{spectoy}
\eea
The oscillating function in the first line of eq \eqref{spectoy} is positive and has a maximal value of $4$,
confirming the results of  \cite{Byrnes:2018txb,Carrilho:2019oqg,Ozsoy:2019lyy}.  The second line
of eq \eqref{spectoy}    exhibits 
this fact  more transparently. In fact, 
  although we work at small scales with sizeable $k$,
they are not really that small: indeed 
we expect
$
 k \,\tau$  to be of order $1$, and ${\Delta \tau_A}/{\tau}\ll1$ for a short duration of the non-slow-roll phase.
 Then, we  expand the first line of eq \eqref{spectoy}  for small value of  $ k \,\tau\,\Delta \tau_A/\tau\,=\, k \, \Delta \tau_A$. In the second line of the formula we learn   that the  maximal  spectral index   is  $n-1\,\simeq\,4$
 in agreement with Fig \ref{fig:complete-profile-cas1}, and with the works
   \cite{Byrnes:2018txb,Carrilho:2019oqg,Ozsoy:2019lyy}.  

\subsubsection{Matching with a second phase of de Sitter expansion}
\label{sec1-match}

After pointing out some universal features of the power spectrum indicated  by our
analytical approach, we make them
more quantitative by matching the system studied so far
 with a second phase of de Sitter expansion from $\tau\,=\,\tau_2$ to $\tau\,=\,0$ \footnote{To simplify our
expressions, we choose units such   that this second phase of pure de Sitter expansion ends at $\tau\,=\,0$. The
end of this second de Sitter phase can correspond to the
end of inflation, but it is not necessarily so, and additional phases of non-slow-roll expansion can occur afterwards, as we discuss  in section \ref{secMNS}.}. 
 For $\tau\le\tau_2$ we have
the solution \eqref{analytic1} for the mode function $\psi_k (\tau)$; for  $\tau\ge\tau_2$ the pump field $z(\tau)$
acquires its `pure-de Sitter' profile $z\,\propto\,1/\tau$ (see eq \eqref{puredS1}) and the corresponding mode function is given by eq \eqref{phidsG}, depending on the time-independent constants ${\cal C}_{1,2}$.
  We can then match solutions \eqref{analytic1} and \eqref{phidsG} and their derivatives 
 through Israel matching conditions 
 at the transition
time $\tau_2$. Assuming that for $\tau\ge \tau_2$ we have pure de Sitter evolution
up to  $\tau=0$ we can identify $$\Delta \tau_B=0-\tau_2$$ as the duration of the second de Sitter 
phase.  Notice the relation 
$|\tau_1|\,=\,\Delta \tau_A +\Delta \tau_B$.

\begin{figure}[h!]
\centering
 \includegraphics[width = 0.7 \textwidth]{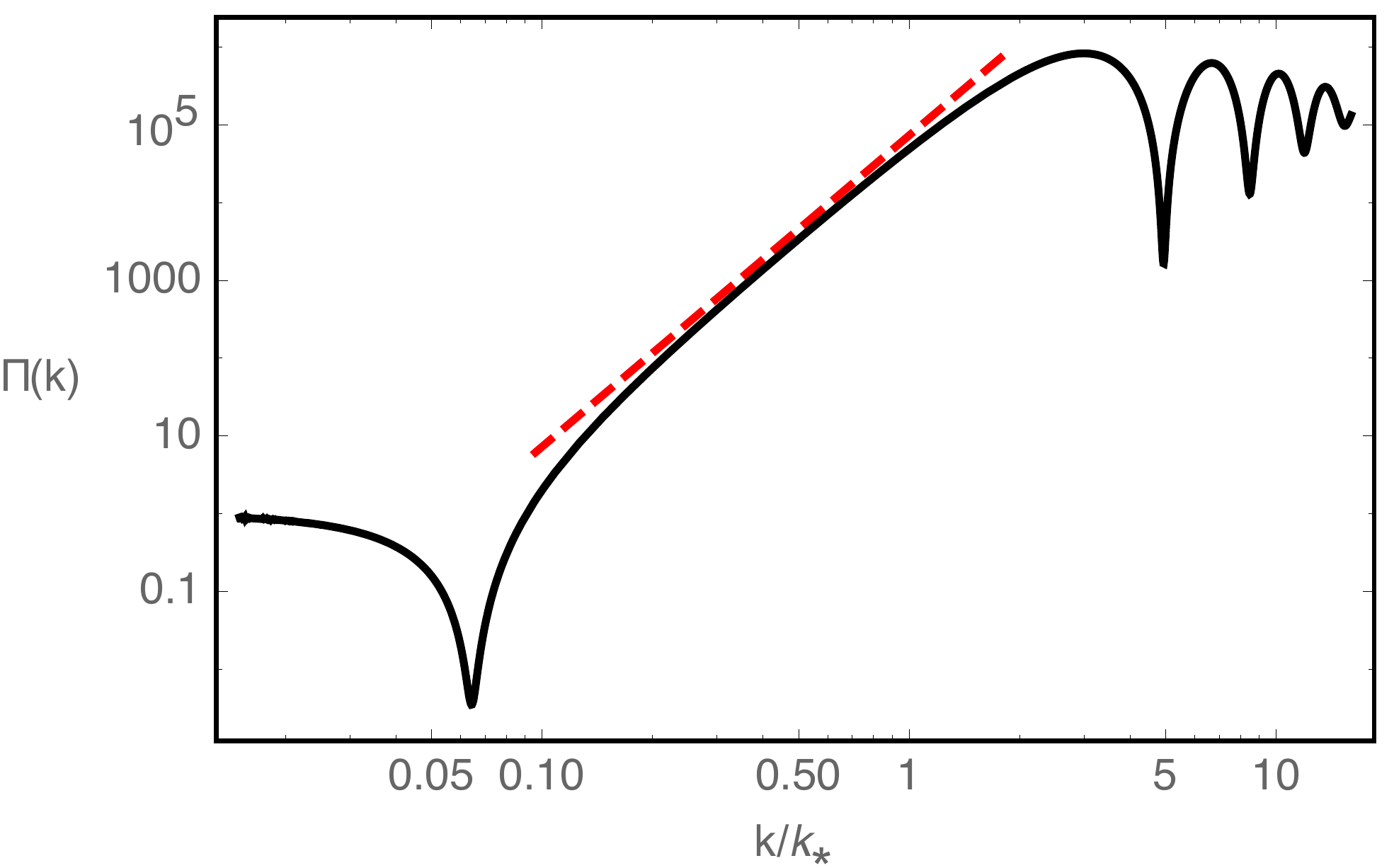}
 \caption{\it Plot of the analytic 
 function $\Pi(k)$  in eq.  \eqref{analydesitt}  versus the dimensionless combination $k/k_\star$, with $k_\star\,=\,\left(\Delta \tau_A+ \Delta \tau_B\right)^{-1}$. We choose
 the representative values $\alpha=4000$, $\Delta \tau_A/\Delta \tau_B\,=\,0.2$. 
 $k_\star$ is
   the horizon-crossing scale 
  for modes leaving the horizon at the onset of the non-slow-roll epoch. 
  Notice that the dip occurs at scales   $k\,<\,k_\star$. 
 The red dashed line has scale-dependent profile
 proportional to $(k/k_\star)^4$. 
   }
 \label{fig:complete-profile2}
\end{figure}

A straightforward computation gives the following values for the constant (but scale-dependent) parameters
${\cal C}_{1,2}$:
\bea
\label{sol1C1}
{\cal C}_1(k)&=&1+\alpha \frac{1-e^{2 i \,k\,\Delta \tau_A}- 2 i   \,k\, \Delta \tau_A (1+ 2 i\, k \,\Delta \tau_B)}{8\,k^2\,\Delta \tau_B^2}\,,
\\
\label{sol1C2}
{\cal C}_2(k)&=&-\alpha\,\frac{e^{2 i \,k\,\Delta \tau_B}}{8
\,k^2\,\Delta \tau_B^2
}\,\left(1-2 i\,k\,(\Delta \tau_A+\Delta \tau_B)-e^{2 i \,k\,\Delta \tau_A}\left(1-2 i\,k\,\Delta \tau_B \right)
\right)\,,
\eea
and the corresponding spectrum $\Pi_k(\tau)$ evaluated at  $\tau\,=\,0$ results simply
\be \label{analydesitt}
\Pi_k(\tau=0)\,=\,\big|{\cal C}_1(k) +{\cal C}_2(k) \big|^2 \,.
\ee
Substituting the  values of ${\cal C}_{1,2}$ the complete expression for $\Pi_k(\tau=0)$ is rather cumbersome,  but it
is straightforward to obtain it combining formulas \eqref{sol1C1}, \eqref{sol1C2} and \eqref{analydesitt}. We avoid to put such long equations in the text, and represent
the result in Fig \ref{fig:complete-profile2} for a representative choice of the parameters.

\subsubsection*{The maximal growth of the spectrum}

The plot \ref{fig:complete-profile2} represents $\Pi_k$ at the end of inflation  as function of the ratio $k/k_\star$, where
we introduce 
the   pivot scale
\be
k_\star\,=\,\frac{1}{\left(\Delta \tau_A+ \Delta \tau_B\right)}\,=\,\frac{1}{|\tau_1|},
\ee 
corresponding to the scale at which the last mode   leave the horizon still during the first quasi-de Sitter phase of expansion. For  scales smaller 
  than $k_\star$ (i.e. $k\ge k_\star$) the modes leave the horizon during the following phase of
  non-slow-roll, 
   or during the second phase of quasi-de Sitter expansion for $\tau>\tau_2$.

 The content of Fig \ref{fig:complete-profile2}  is qualitatively similar to  what  discussed in section \ref{sec1pssv}: it presents  a dip in the spectrum at relatively large scales (that left the horizon well earlier than the onset of the non-slow-roll epoch) as well as a maximal slope $n-1\,=\,4$ for the growth of the spectrum  after the dip occurs. At very small scales (large $k/k_\star$) the spectrum asymptotes on average to a constant, and the `oscillatory' behavior is due to our instantaneous matching 
of the mode function to the pure de Sitter mode solution. It is easy to take the asymptotic limit of the function 
\eqref{analydesitt} at very small scales. We introduce the dimensionless quantity
\be\label{defeA}
\sigma_{A}\,=\,\frac{\Delta \tau_{A}}{\Delta \tau_A+\Delta \tau_B}\,=\,\frac{\Delta \tau_{A}}{|\tau_1|}\,,
\ee
that corresponds to the small parameter parameterizing the short duration of the non-slow-roll phase (see discussion
around eq. \eqref{smdtr}).
 Using this parameter find  the expression
\bea
\label{PIMAXd}
\Pi_{\max}&=&\left(1+\frac{\alpha}{2}\,\frac{\Delta \tau_A}{\Delta \tau_B} \right)^2\,=\,
\left(1+\frac{\alpha}{2}\,\frac{\sigma_A}{1-\sigma_A} \right)^2\,,
\eea
for the  value of the spectrum at very small scales.  This expression
  has the expected properties for such asymptotic limit: it diverges when $\Delta \tau_B\,\to\,0$ (equivalently $\sigma_A\to1$),  reflecting
the fact that in absence of the last phase of de Sitter expansion the spectrum grows unbounded. In the limit of large
$\alpha$, keeping $\sigma_A$ fixed, the value of $\Pi_{\max}$   scales as $\alpha^2$, showing that the parameter $\alpha$  controls the amplitude
of the spectrum. 
 On the other hand, comparing with the arguments developed in section \ref{sec-anfor} (see
 in particular the discussion after eq \eqref{smdtr}), we need to be aware that
 too large values of the quantity $\alpha\,\sigma_A$ might be incompatible
 with the truncation of our perturbative expansion, unless cancelations occur that make small the
 higher-order Taylor contributions. This is a model-dependent issue that we do not discuss in this work,
 leaving it for future investigations (but see below for a possible resolution of this problem).

Two  important points are worth emphasizing:
\begin{itemize}
\item
 In plotting Fig \ref{fig:complete-profile2} we choose a  large value of $\alpha$, in order to obtain a 
  spectrum growth of order $\sim 10^5$ from large to small scales: in our discussion here $\alpha$
is a tunable parameter that we treat phenomenologically,  and we do not discuss  possible scenarios giving so large values. On the other hand,  
too large values of $\alpha$ can be not easy to realize in concrete   model building,  or in tensions with the 
reliability of the perturbative scheme of section \ref{sec-anfor}.  In theses cases, we can
   break down the time-interval $\Delta \tau_A$ of non-slow-roll evolution
in smaller `time-bins' --  each characterized by a smaller $\alpha$ --  that combined together
accumulate the necessary large growth of the spectrum. We briefly explore this possibility in section
 \ref{secMNS}, showing that indeed the resulting profile of the 
 spectrum is similar to  Fig \ref{fig:complete-profile2}.

\item  In the example we study the amplitude of the spectrum tends to a constant at small scales after reaching the peak, which
is of the same order of the size of the peak (this is why we call it $\Pi_{\rm max}$ in eq \eqref{PIMAXd}). In other studies, the spectrum tends to decrease with $k/k_\star$
after the peak have been reached. This depends on the (model-dependent)  dynamics  and properties
 of the  last stages of the inflationary expansion after the end of the non-slow-roll phase. In what follows, we  
 call   $\Pi_{\rm max}$ the size  of the  enhancement  in the amplitude of  the spectrum at around the peak position. 
\end{itemize}

\subsubsection*{General  properties of the dip}

Working on a large $\alpha$ limit, we can study the position and amplitude of the dip making a large-$\alpha$ expansion
of the spectrum expressed in terms of the variable $x$ as
\be\label{ans2x}
\frac{k}{k_\star}\,\equiv\,\frac{x}{\sqrt{\alpha}}\,,
\ee
analogously to the arguments  explained around eq \eqref{anstA}. 
 This Ansatz is motivated by the fact that the dip position occurs at relatively large scales, and 
 it is then expected to be controlled by some inverse powers of $\alpha$, the only large parameter
 available. This Ansatz turns 
 to be correct: in the large $\alpha$ limit,
the resulting spectrum 
 is expressed as a part that does not depend on $\alpha$, plus subleading  contributions suppressed by powers of $\alpha$ -- exactly analogously to what found in eq \eqref{PiXexp}. The part that does not depend on $\alpha$ vanishes at a scale
\be
\frac{k_{\rm dip}}{k_{\star}}\,=\,\frac{\sqrt{3}}{\sqrt{1-\sigma_A} \sqrt{2 \sigma_A+1}}\,\frac{1}{({{\Pi^{1/2}_{\rm max}} -1})^{1/2}}\,,
\ee
indicating the position of the dip, with $\Pi_{\max}$ given in eq \eqref{PIMAXd} and $\sigma_A$
in eq \eqref{defeA}. If $\Pi_{\rm max}$ is large, and we take $\sigma_A\ll1$  -- consistently  with the fact that we
take a short duration for the non-slow-roll epoch -- the previous relation simplifies giving
\be
\label{kdipapp}
\frac{k_{\rm dip}}{k_{\star}}\,\simeq\,\sqrt{\frac32}\,\frac{1}{\Pi_{\rm max}^{1/4}}\,,
\ee
 providing us with a simple,   universal relation for the position of the dip as a function of $\Pi_{\rm max}$,
the maximal growth of the spectrum. 
 This relation applies to scenarios characterized by a single brief phase of drastic violation from slow-roll
 expansion.
 Notice that $k_{\rm dip}$ is generally well smaller than $k_\star$, the
scale when non-slow-roll starts, and at leading order is independent on the actual duration of the short
non-slow-roll phase.

\begin{figure}[h!]
\centering
 \includegraphics[width = 0.7 \textwidth]{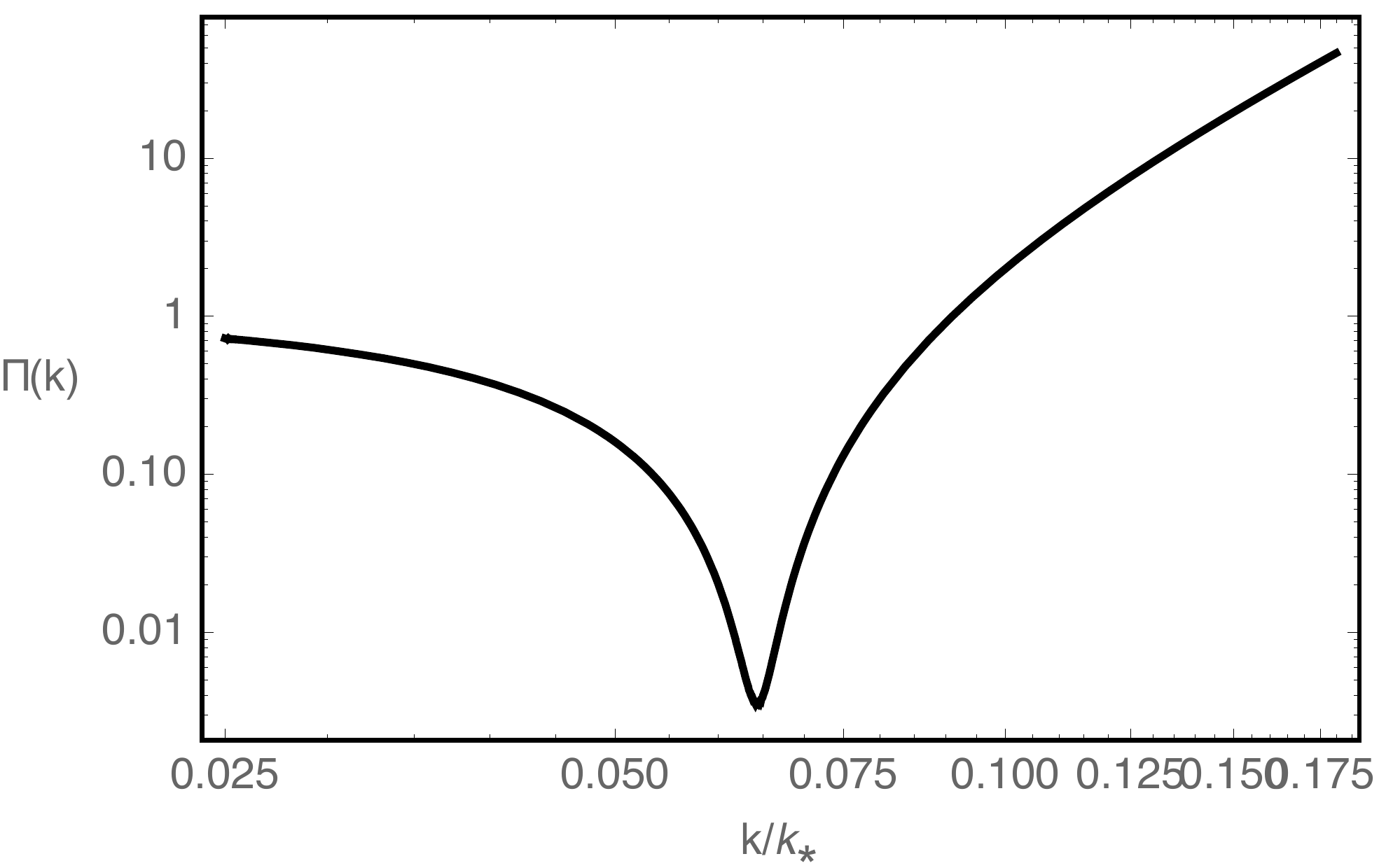}
  \caption{\it Zoom on the dip of Fig \ref{fig:complete-profile2}.  We choose
 the representative values $\alpha=4000$, $\Delta \tau_A/\Delta \tau_B\,=\,0.2$ 
  for the parameters that enter 
  in   eq.  \eqref{defPI}. Our formulas  \eqref{kdipapp}, \eqref{kdipapp2} give $k/k_*=0.064$ while $\Pi(k_{\rm dip})=0.0035$, in 
  agreement with the plot.
  }
 \label{fig:complete-profile2dip}
\end{figure}

Importantly, this relation 
  connects distinct
 quantities in a way that is independent on the values of the model parameters (as $\alpha$)
 and only depends on physical properties of the spectrum. 
 The depth of the dip is also straightforward to determine: calling $\Pi_{\rm min}\,=\,\Pi_{k\,=\,k_{\rm dip}} (0)$, we find
\bea\label{kdipapp2}
\Pi_{\rm min}&=&\frac{12}{(1-\sigma_A) ( \sigma_A+2)^3}\,\frac{1}{({{\Pi^{1/2}_{\rm max}} -1})}\,,
\\
&\simeq&{\frac32}\,\frac{1}{\Pi_{\rm max}^{1/2}}\,=\,\frac{k^2_{\rm dip}}{k_{\star}^2}\,.
\eea
Then also the depth of the dip is related with the value of the maximal growth of the spectrum through the previous
 universal relation. Fig \ref{fig:complete-profile2dip} shows that our analytic formulas  \eqref{kdipapp}, \eqref{kdipapp2} work very well in determining the
 properties
 of the dip for the example represented in Fig \ref{fig:complete-profile2}.

The information we derived, besides being theoretically interesting, is also relevant for phenomenology. 
Starting from $k_{\rm dip}$ the spectrum then starts a steady growth towards its maximum. The position  of the dip is important being  the  characteristic scale at which the spectrum profile starts to considerably deviate from slow-roll
predictions, and we need to ensure its
profile is consistent with large-scale constraints from CMB  and spectral distortions
 \cite{Bugaev:2010bb,Nakama:2017xvq,Byrnes:2018txb,Sato-Polito:2019hws,Unal:2020mts}. 

\smallskip

Our  relations   \eqref{kdipapp}, \eqref{kdipapp2}  are obtained under the hypothesis that 
before  the non-slow-roll epoch the universe is
in a de Sitter phase, with a mode function for perturbations matching a Bunch-Davies vacuum at early times. We derived this formula in a limit of short non-slow-roll phase, truncating the Taylor series as explained in Section \ref{sec-anfor}. 
We
connect our non-slow-roll phase abruptly with a phase of de Sitter expansion towards the end of inflation. 
 Despite these many assumptions, these formulas match well analogous  formulas determined in the literature using other analytic methods. For example, the authors of \cite{Byrnes:2018txb} study the growth of the spectrum in a set-up including a phase of ultra-slow-roll  inflation, making the hypothesis that in the epoch of slow-roll violation the slow-roll parameter $\epsilon_2$ can be large but constant, and analysing the system using
   Israel matching conditions.  They find for the position of the dip an expression ($N_{USR}$ being
   the number of e-folds of ultra-slow-roll expansion)
   \be \label{dimcoby}
   \frac{k}{k_\star}\,\simeq\,\sqrt{\frac{5}{4}}\,\left( \exp{\left[6\,N_{USR}\right]}\right)^{-1/4}\,\propto 
   \,\sqrt{\frac{5}{4}}\,\Pi_{\rm max}^{-1/4}\,,
   \ee
   in good overall agreement with our formula \eqref{kdipapp}.

 Given these preliminary results, 
   it would be interesting to study in more general terms how precise these formulas can be in fully realistic models with non-slow-roll behaviour, possibly including corrections associated with a longer duration of the non-slow-roll phase, and a more realistic matching to the final phase of slow-roll expansion. 
   We leave these investigations to future work. In section \ref{secMNS} we instead  study how these results are modified in the presence of multiple short phases of non-slow-roll evolution.  

 \subsubsection*{The slope of the spectrum}

Fig \ref{fig:complete-profile-cas1} shows that the maximal slope the spectrum is $n-1\,=\,4$, as found in \cite{Byrnes:2018txb}. 
The work \cite{Carrilho:2019oqg} found an interesting physical explanation of this fact in terms of the behaviour
of the mode functions during epochs of non-slow-roll, under the hypothesis that the (possibly large) slow-roll
parameters are constant. 
Our approach provides an alternative
 analytic understanding of this behaviour without any assumption on the slow-roll parameter
 behaviour,   using the same arguments
we developed  in section \ref{sec1pssv}. We focus in a region of scales  $k$ well after the dip, when the large-$\alpha$ contributions to the spectral index do not risk to be compensated by extremely  small coefficients of $k/k_\star$. 
In a region of  $0.1\,<\,k/k_\star\,<\,1$ we can expand the spectral index for large $\alpha$, small $k/k_\star$, and 
small $\sigma_A$, 
finding
\be
n-1\,=\,4-\frac{k^2}{30\,k_\star^2} \left(12-12 \sigma_A+7 \sigma_A^2\right)+\dots
\ee
confirming the maximal slope $n-1\,=\,4$. 

\smallskip

It is  interesting   to ask whether we can overcome this threshold within single-field
inflation. The work \cite{Carrilho:2019oqg} (see also \cite{Ozsoy:2019lyy}) 
shown that if a phase of non-attractor inflation is preceded  by a prolonged epoch of non-slow-roll evolution, then
 the growth of the spectrum  can increase up a  power $k^5\,\ln k$, indicating that the slope of the spectrum during non-slow-roll epoch
 retains some sort of `memory' of  previous phases of evolution of the pump field $z(\tau)$. The 
 finding of  \cite{Carrilho:2019oqg}  was obtained by making a perturbative expansion of solutions
 of mode functions in regime of constant (possibly large) slow-roll parameters.
 
 We wish to elaborate more on this topic in the next section using our approach. We will learn
 that indeed, for certain single-field systems,  the spectrum can grow steeper
    than what has been
 found so far.

\subsection{Multiple non-slow-roll epochs}
\label{secMNS}

In this section  we
investigate whether multiple short phases of non-slow-roll evolution, nested within epochs
of de Sitter expansion, can change the results of the previous section.  In particular, we examine
whether such scenario can lead to a more pronounced  slope for the spectrum, that overcomes
the bound $n-1=4$ for the spectral index. To analyse this question we make
use of the methods developed in the previous sections, and in Appendix \ref{App_comparison}  we  
compare our results with other approaches. 

\smallskip

The possibility of many phases 
of evolution that deviate from slow-roll during the inflationary period might be realized in scenarios with
several local extrema with  inflection points in the inflationary potential, as in axion inflation models which include
subleading non-perturbative corrections
  (see e.g. \cite{Parameswaran:2016qqq,Ozsoy:2018flq,Bizet:2016paj,Kobayashi:2017jeb,Ozsoy:2020ccy}), in models based
 on axion monodromy \cite{Silverstein:2008sg,McAllister:2008hb}, 
in scenarios 
where the inflationary potential has several jumps and discontinuity in its first derivatives 
 \cite{Starobinsky:1992ts,Adams:1997de}, in rollercoaster cosmology \cite{DAmico:2020euu}. Besides inflation, scalar potentials
 with many extrema have been considered in scenarios  proposing a cosmological relaxation of 
   the cosmological constant \cite{Abbott:1984qf} or the electroweak scale \cite{Graham:2015cka}.
   In any case, regardless of 
concrete constructions and models,  we aim to  study here some phenomenological consequences
of this possibility. 

As mentioned above, one of our motivations is
 the work \cite{Carrilho:2019oqg} (see also \cite{Ozsoy:2019lyy})  that 
suggests that phases of non-slow-roll preceding an epoch of non-attractor can change the slope
of the spectrum. 
Intuitively, a more pronounced slope of the spectrum can be heuristically understood from our formulas.
The results of section \ref{sec1-match} indicate that the constant quantities ${\cal C}_{1,2}$ characterising
the de Sitter mode function \eqref{phidsG} in the  de Sitter phase following the non-slow-roll epoch 
store a rich $k$-dependence, that leads to the $k$-dependent profile of the complete spectrum
 of fluctuations when evaluated towards the end of inflation -- see indeed eqs \eqref{sol1C1}, \eqref{sol1C2}. 
  If -- before ending inflation -- we enter in 
 a new, second phase of non-slow-roll evolution, we can use
  the de Sitter mode function with coefficients  ${\cal C}_{1,2}$
 of eqs \eqref{sol1C1}, \eqref{sol1C2}
   as a basis for determining the mode function during the non-slow-roll epoch. 
  This fact can drastically change the resulting profile of the final spectrum, which accumulates
  both the $k$-dependence associated with the second non-slow-roll phase, as well as the $k$-dependence
  already stored in the `initial conditions' governing  ${\cal C}_{1,2}$. 
  The final slope of the spectrum
can then retain 
memory of all phases of non-slow-roll expansion.

\begin{figure}[h!]
\centering
 \includegraphics[width = 0.7 \textwidth]{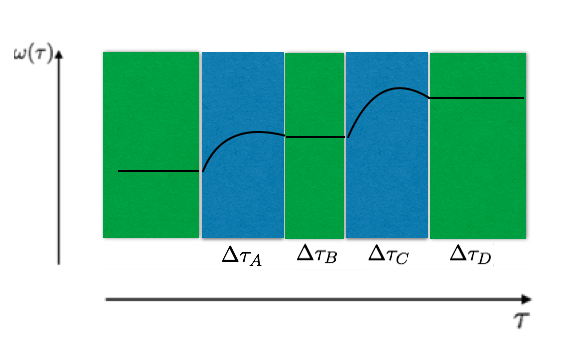}
 \caption{\it  Pictorial representation of the distinct phases of evolution, in a system  
 extending the example of Fig \ref{fig:artistic} to the case of multiple non-slow-roll epochs. Recall that
 the function $\omega(\tau)$ controls departures of the pump field $z(\tau)$ from its de Sitter profile (see eq \eqref{zgen}).
   Two
 phases of non-slow-roll evolution (in blue) are interrupted by  a de Sitter epoch (in green). De Sitter expansion also characterises  the epochs preceding and following the non-slow-roll periods. }
 \label{fig:arte2}
\end{figure}

As a concrete realization, we consider a multiple-phase system consisting on the following epochs (recall that we work with negative values of $\tau$):
\begin{itemize}
\item[0)] {\bf{$-\infty\le\tau\le\tau_1$}}: standard de Sitter expansion, with de Sitter mode function \eqref{phidsG}
 and  Bunch-Davies
vacuum (${\cal C}_1\,=\,1$, ${\cal C}_2\,=\,0$).
\item[1)] $\tau_1\le\tau\le \tau_2$: first phase of non-slow-roll, described by a mode function \eqref{analytic1} with characteristic parameter $\alpha$. The  duration of this short phase is $\Delta \tau_A\,=\,\tau_2-\tau_1$. 
\item[2)] $\tau_2\le\tau\le\tau_3$: an intermediate phase of de Sitter expansion,  with de Sitter mode function \eqref{phidsG}, and parameters ${\cal C}_{1,2}$ determined by matching conditions as in eqs \eqref{sol1C1}, \eqref{sol1C2}.
 The  duration of this phase is $\Delta \tau_B\,=\,\tau_3-\tau_2$. 
\item[3)] $\tau_3\le\tau\le\tau_4$: second phase of non-slow-roll, described by a mode function \eqref{analytic1a} with characteristic parameter $\beta$. The  duration of this short phase is $\Delta \tau_C\,=\,\tau_4-\tau_3$. 
\item[4)] $\tau_4\le\tau\le0$: last phase of de Sitter expansion,  with de Sitter mode function \eqref{phidsG}, and parameters ${\cal C}_{1,2}$ determined by matching conditions. This phase concludes inflation. 
 The  duration of this epoch is $\Delta \tau_D\,=\,-\tau_4$. 
\end{itemize}

\begin{figure}[h!]
\centering
 \includegraphics[width = 0.47 \textwidth]{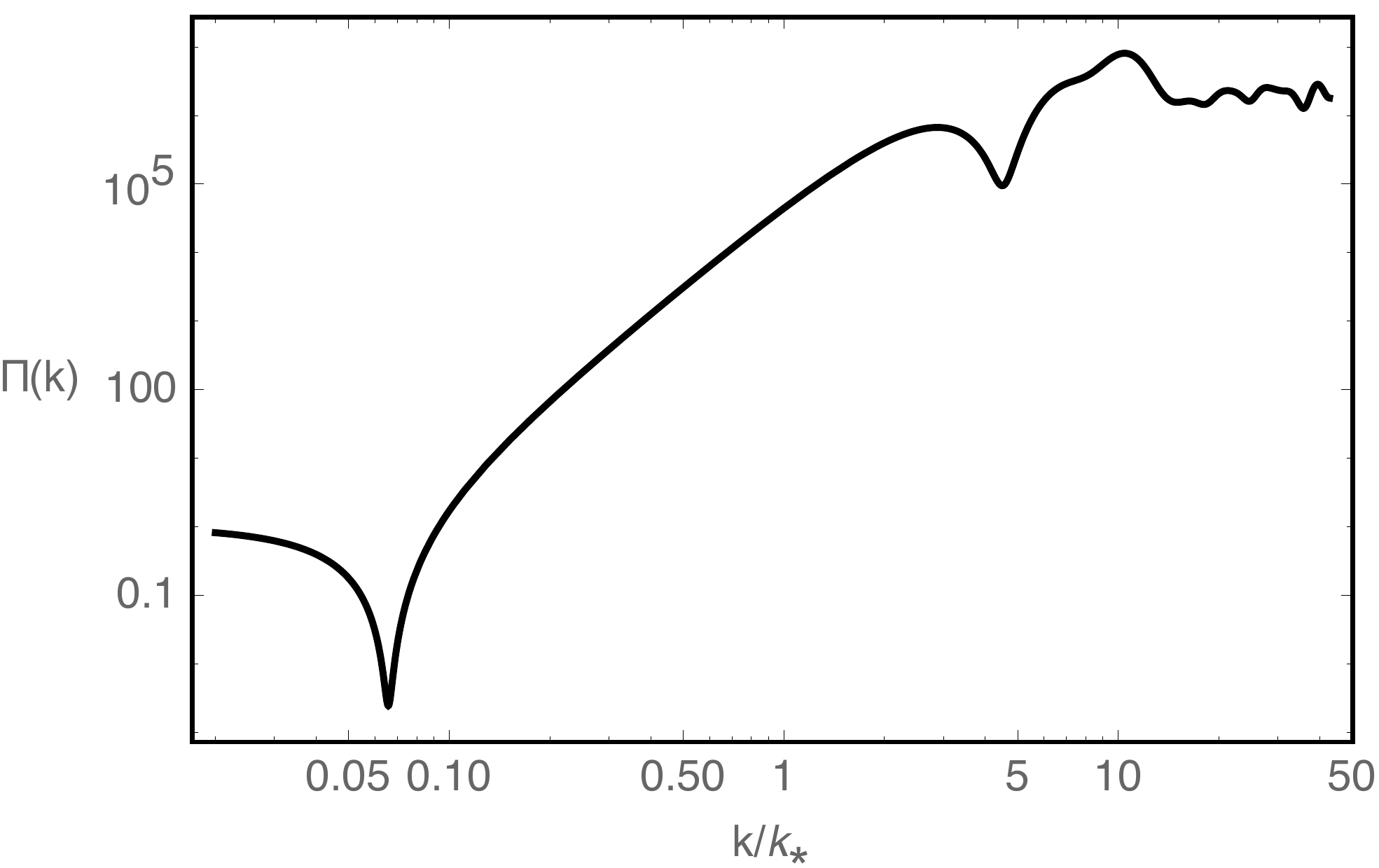}
  \includegraphics[width = 0.47 \textwidth]{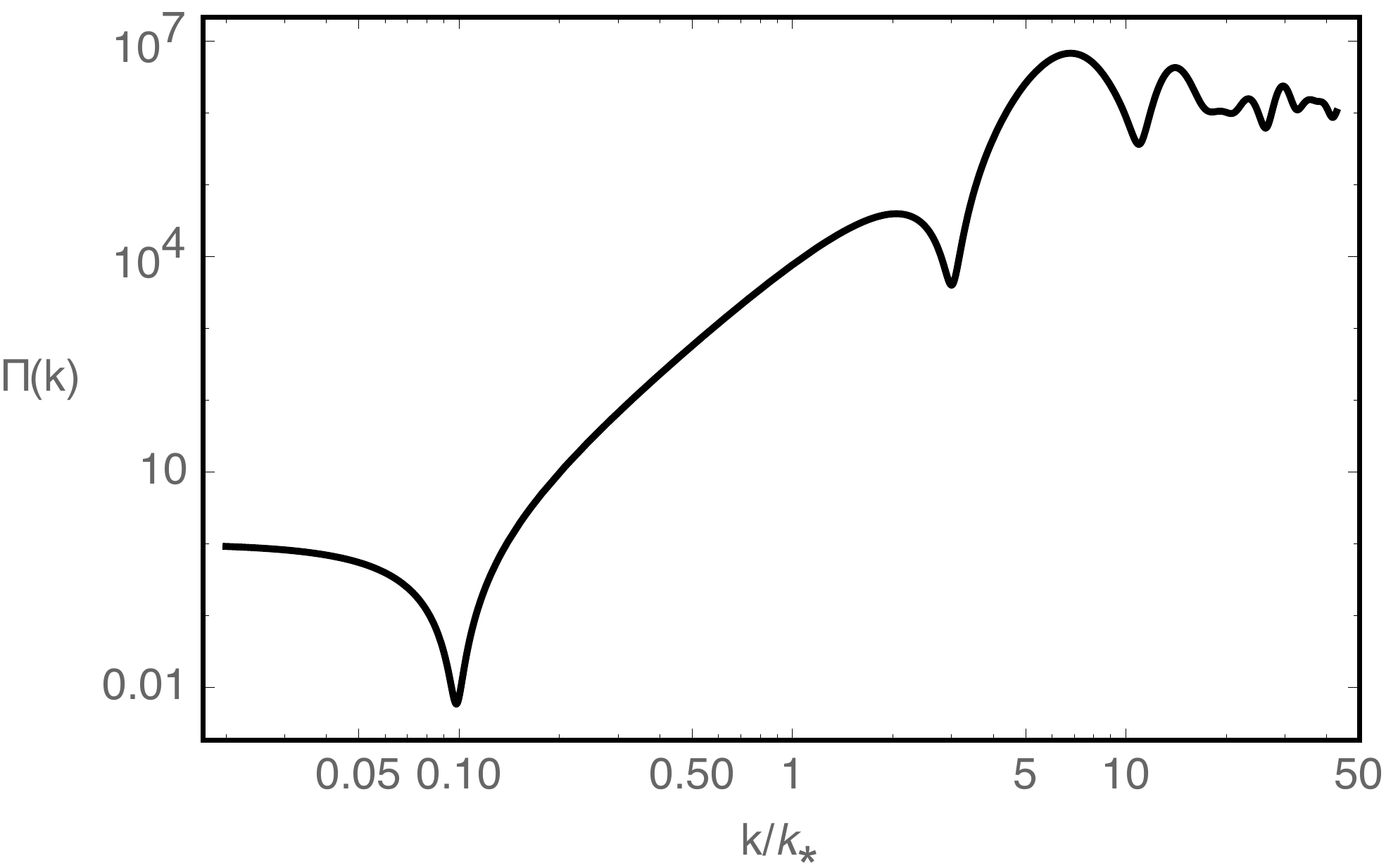}
 \includegraphics[width = 0.7 \textwidth]{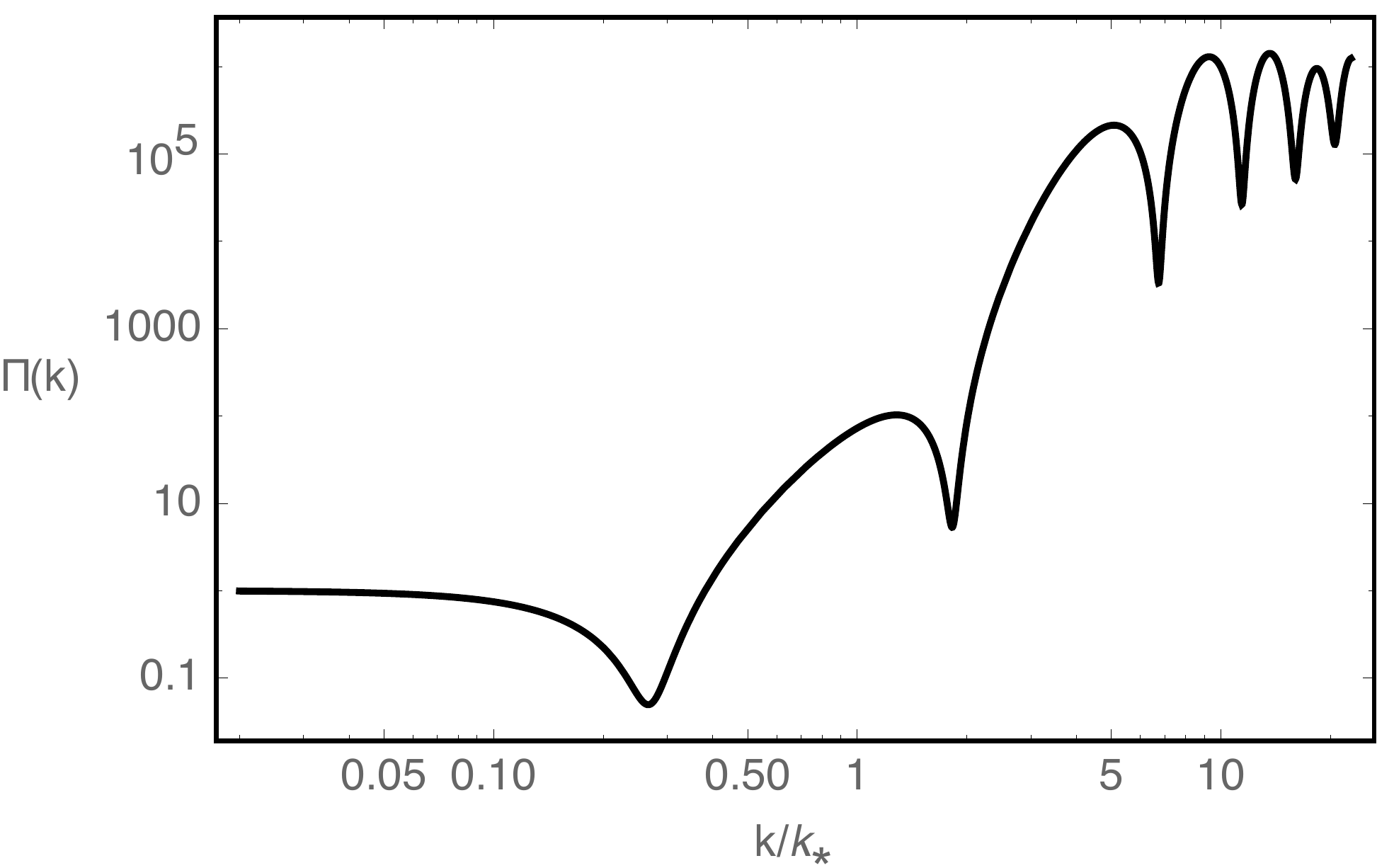}
  \caption{\it Three examples of profile of the fluctuation spectrum for the system discussed
 in this section.  
The three plots differ for the duration of the intermediate de Sitter phase
 between the two phases of non-slow-roll evolution. We choose the parameters $\alpha\,=\,250$, and $\beta\,=\,200$. The duration of the non-slow-roll phases is $\Delta \tau_A\,=\,\Delta \tau_C\,=\,0.2$, and the final de Sitter
 phase lasts $\Delta \tau_D=0.7$. For what respect the intermediate de Sitter phase we choose respectively:  \bf{Top Left} $\Delta \tau_B\,=\,0.2$; \bf{Top Right} $\Delta \tau_B\,=\,0.5$; \bf{Bottom} $\Delta \tau_B\,=\,2.0$.  }
 \label{fig:complete-profile3ab}
\end{figure}

Our formalism
  allows us to analytically obtain  the mode functions and the power spectrum at the end of inflation, at least
with the help of some computer program for manipulating algebraic expressions. We do not write the 
cumbersome final expression for the spectrum at the end of inflation; instead we represent it in Fig \ref{fig:complete-profile3ab} for distinct durations of the
intermediate de Sitter phase $\Delta \tau_B$. In representing the plots, we choose as   pivot scale 
$$k_\star\,=\,\left( \Delta \tau_A+ \Delta \tau_B+ \Delta \tau_C+ \Delta \tau_D\right)^{-1}\,=\,|\tau_1|^{-1}\,,$$ 
corresponding to the scale where modes start leaving the horizon when the first non-slow-roll phase begins. 

\smallskip

\smallskip
Before starting any quantitative analysis, we can point out interesting features of these plots:
 \begin{itemize}
\item 
 First,  the upper left plot of Fig \ref{fig:complete-profile3ab}
 shows
 a spectrum profile that is qualitatively  similar to the plots we found in the previous section, see e.g. Fig
\ref{fig:complete-profile2}. This plot corresponds to a short duration of the intermediate de Sitter phase between the two non-slow epochs. The spectrum is enhanced by a factor of  order $\sim 10^5$ from large to small scales,  with  values of the parameters $\alpha$, $\beta$ that are one order of magnitude smaller than the ones we considered in section \ref{sec1pssv}. This suggests 
that we might study inflationary systems containing longer phases of non-slow-roll evolution by `breaking down' such
non-slow-roll phases in  small time bins, that can be then joined together to get the final result.  In each bin  the pump field profile can be faithfully characterised with few parameters, and the perturbative expansion developed in section \ref{sec-anfor} could be applied more reliably.
 We leave an analysis of this possibility to future work.
\item The behaviour of the spectrum profile in the plot at the bottom of  Fig \ref{fig:complete-profile3ab} is instead
 different with respect to what we have found so far. It corresponds
to a system with a prolonged phase of de Sitter evolution, nested between the two epochs of non-slow-roll
expansion. After the initial dip, the spectrum of  Fig \ref{fig:complete-profile3ab} grows few orders
of magnitude, before meeting another intermediate dip. After the second dip, the spectrum encounters
a more enhanced phase of growth, with  an amplification by many  orders of magnitude,
which is  characterized by a  more pronounced slope. 
 In what follows  we  
  analyse this case, since it can offer opportunities
to increase the slope of the perturbation spectrum in single-field inflation.  
\end{itemize}

\subsubsection*{The maximal growth of the power spectrum}

A straightforward computation  analog to the ones we pursued in the previous section
  leads to the following small-scale (large $k/k_\star$) limit for the spectrum evaluated at the end of inflation

\be
\Pi_{\max}\,=\,
\left( 
\frac{\left(\beta \Delta \tau_C+2 \Delta \tau_D  \right)
\left( 
\alpha\,\Delta \tau_A+2 ( \Delta \tau_B+ \Delta \tau_C+ \Delta \tau_D) \right)
}{4\,\Delta \tau_D\, ( \Delta \tau_B+ \Delta \tau_C+ \Delta \tau_D) }
\right)^2\,.
\ee
This quantity represents the maximal growth of the spectrum: setting  $\Delta \tau_B=\Delta \tau_C\,=\,0$
one finds the same formula of eq \eqref{PIMAXd} since we reduce to the case of   section \ref{sec1pssv}. In the limit of large
$\alpha$, $\beta$ it simplifies to the expression
\be
\Pi_{\max}\,=\,\frac{\alpha^2 \beta^2}{16} \left(\,\frac{\Delta \tau_A\,\Delta \tau_B}{\Delta \tau_D \left( 
\Delta \tau_B+
\Delta \tau_C+
\Delta \tau_D
\right)} \right)^2\,.
\ee
The amplitude of the spectrum is now given by the product of the squares of both parameters
$\alpha$ and $\beta$. This is why we can amplify the amplitude of the spectrum by a factor $\sim 10^5$
with smaller values of the phenomenological parameters $\alpha$, $\beta$
with respect to section \ref{sec1pssv}. 

\subsubsection*{Properties of the first dip}

We can determine the position of the first, pronounced dip in plots \ref{fig:complete-profile3ab}  by 
rescaling variables as done in the previous section around eq \eqref{ans2x}:

\be\label{ans2x2}
\frac{k}{k_\star}\,\equiv\,\frac{x}{\sqrt{\alpha}\, \sqrt{\beta}}\,,
\ee
 and take the large $\alpha$, $\beta$
limit of the spectrum.  We define the parameters
\bea
\sigma_A&=&k_\star\,\Delta \tau_A\hskip0.5cm,\hskip0.5cm\Sigma_B\,=\,k_\star\,\Delta \tau_B
\\
\sigma_C&=&k_\star\,\Delta \tau_C\hskip0.5cm,\hskip0.5cm\Sigma_D\,=\,k_\star\,\Delta \tau_D
\eea
following the criterium that parameters  denoted with small letters $\sigma_{A,\,C}$ parameterize the short
duration of non-slow-roll epochs, while the ones denoted with large  letters $\Sigma_{B,\,D}$ parameterize the arbitrary duration of de Sitter epochs.

We
find the following expression for the position of the dip: 
\be
\frac{k_{\rm dip}}{k_\star}\,\simeq\,\left[ \frac{3 \left(\Sigma_B+\Sigma_D\right)^2}{4\,\sigma_C\,\Sigma_D^2\,
 \left(\Sigma_B+2 \Sigma_D\right)
} \right]^{1/2}\,\frac{1}{\Pi_{\rm max}^{1/4}}\,.
\ee
We derived this expression in the limit of large $\Pi_{\rm max}$, and of short duration of the non-slow-roll phases (small $\sigma_{A,\,C}$). 
This formula is slightly less accurate in determining the position of the dip with respect to the formulas
we found in the previous section, due to the fact that in this more complex system the dip width is larger
than in the single non-slow-roll phase. But in studying various examples we find that it is accurate up to a factor
of order one in determining $k_{\rm dip}$.  What is interesting is that the dip position in this case is  tunable: the
position again scales as the inverse of the fourth root of   $\Pi_{\rm max}$, but it also depends on the duration of
the  non-attractor epochs as well on the intermediate de Sitter phase.

\subsubsection*{The slope  of the spectrum towards small scales}

\begin{figure}[h!]
\centering
 \includegraphics[width = 0.495 \textwidth]{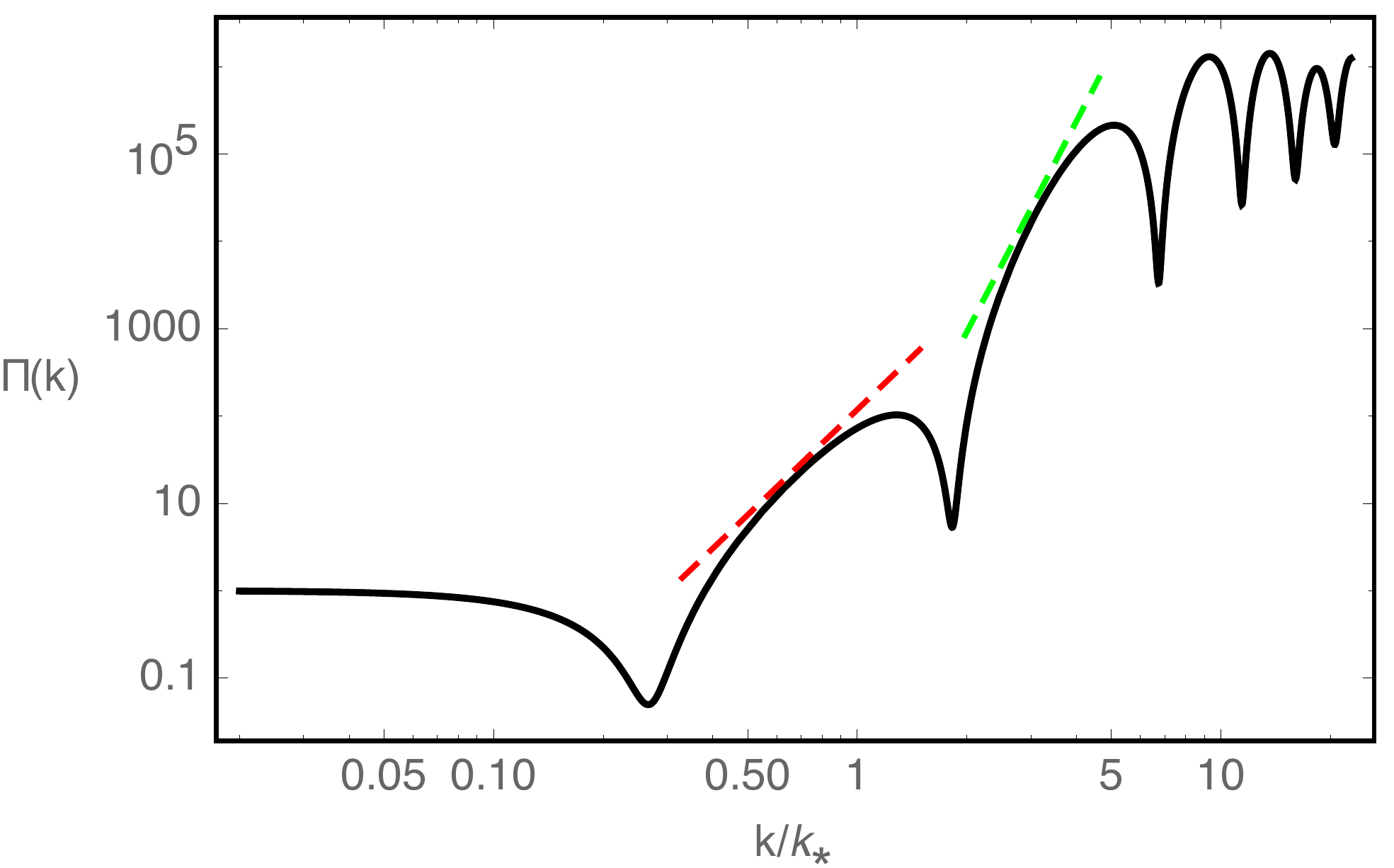}
  \includegraphics[width = 0.495 \textwidth]{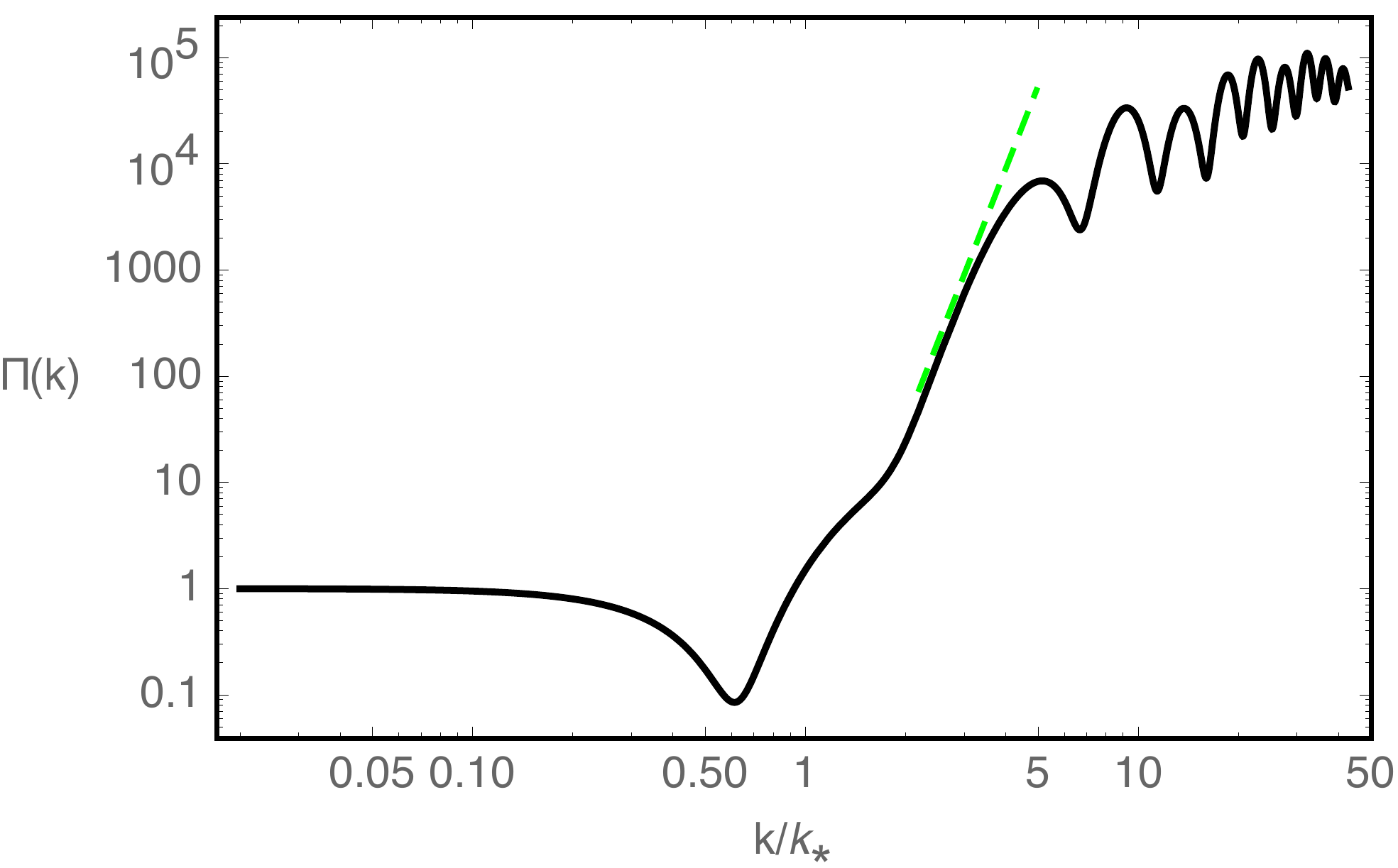}
 \caption{\it Two examples of spectra computed with our analytic formulas. The green dashed line
 has a profile $k^8$, suggesting that the example of this section can in principle enhance
 the spectrum. The red dashed line has a profile $k^4$.   {\bf Left}: We choose the parameters $\alpha=250$, $\beta=200$ and
 $\Delta \tau_A\,=\,0.3$,  $\Delta \tau_B\,=\,2.0$,  $\Delta \tau_C\,=\,0.3$, $\Delta \tau_D\,=\,0.7$. {\bf Right}:
  $\alpha=30$, $\beta=1000$ and
 $\Delta \tau_A\,=\,0.3$,  $\Delta \tau_B\,=\,3.5$,  $\Delta \tau_C\,=\,0.3$, $\Delta \tau_D\,=\,1.4$.
 }
 \label{fig:complete-profile3ac}
\end{figure}

An analysis of the slope of the spectrum  is particularly interesting. As mentioned above, the spectral index 
can turn out to be  larger in the phase of rapid growth of the spectrum, after the second intermediate dip
occurs.
 Fig \ref{fig:complete-profile3ac} represents the slope of the spectrum in two situations.

The green dashed curves have $n-1\,=\,8$, suggesting that this can be the maximal slope of the
spectrum towards small scales, in the region of more enhanced growth. This overcomes considerably the $n-1=4$ limit of \cite{Byrnes:2018txb}, that we also found
in the previous section focussing on a single phase of non-slow-roll expansion.

\smallskip

This behaviour can be  analytically understood by means of our formulas, exploiting
arguments we introduced in section \ref{sec1-match} for studying the slope of the spectrum. We compute the
spectral index associated with the spectrum in this system, and focus on the limit of large $\alpha$,
$\beta$ in a range of scales  well after the first dip. Formulas are complex, but making appropriate
expansions in regimes of small parameters they simplify considerably and are amenable of 
analytical investigation. First, we do an expansion of $n-1$ for small values of $\sigma_A$ and $\sigma_C$
and retain only the leading terms in the expansion, since we are interested to the limit of  short
duration of non-slow-roll epochs. We are left with the parameters $\Sigma_B$, $\Sigma_D$
 and $k/k_\star$. From Fig \ref{fig:complete-profile3ab}, we know that having a sizeable  $\Sigma_B$ is essential
 for making manifest the enhanced slope of the spectrum at small scales. Instead, having a small $\Sigma_D$ 
 allows for further amplify the spectrum at small scales: we can then use $\Sigma_D$ as additional small quantity for   our arguments.
 
 Namely,  already working in a  leading order expansion in  the small quantities $\sigma_A$ and $\sigma_C$, 
 we then make two different additional expansions of the spectral index  $n-1$  in the two independent   quantities  $k/k_\star$ and  $\Sigma_D$. In the first expansion we impose the  hierarchy $k^2/k^2_\star \,<\,\Sigma_D$, which corresponds to
   relatively large scales between the two dips in the left panel of Fig \ref{fig:complete-profile3ac}. The spectral index results
  \be
n-1\,=\,4-\frac{2\,\Sigma_B^3\,k^2}{\Sigma_D\,k_\star^2}+\dots
\ee
where dots indicate subleading contributions in this regime. This formula matches with the red dashed line in the left panel of Fig \ref{fig:complete-profile3ac}. 
 In the second expansion we consider  smaller scales after the second dip,  and impose the hierarchy $k^2/k^2_\star \,>\,\Sigma_D$. 
We then find
\be
n-1\,=\,8-\frac{2 \,\Sigma_B\,k^2}{5\,k_\star^2}+\frac{8\,\Sigma_D\, k_\star^2}{\Sigma_B^3\,k^2}+\dots
\ee
in agreement with  the green dashed line in the left plot of Fig \ref{fig:complete-profile3ac}. 
  These findings suggest that a succession of non-slow-roll phases in single-field inflation can much enhance
the slope of the spectrum in single field inflation. In Appendix \ref{App_comparison} we support these findings by discussing
the same system from the perspective of a succession of de Sitter and ultra-slow-roll phases, following the methods of \cite{Byrnes:2018txb}. 
 Of course, we might expect that including three or
more distinct separate phases of non-slow-roll evolution one can enhance the slope even more --
we leave to future work
an analysis of this fact and the  consequences of these findings.

\subsubsection*{The induced gravitational wave spectrum}

We  conclude briefly discussing the
 possibility that the enhanced slope found here refers to the spectrum of curvature fluctuations, that
 being amplified can generate at second order a stochastic gravitational wave (GW) background. The  scale-dependent profile of the latter might have an amplitude to be sufficiently large to be directly detectable with GW
  experiments. The formalism to study this process was first developed in \cite{Mollerach:2003nq,Ananda:2006af,Baumann:2007zm}, and recently reconsidered or extended in various works (see
  for example \cite{Saito:2008jc,Nakama:2016enz,Garcia-Bellido:2016dkw,Cai:2019cdl,Espinosa:2018eve}), also in view
 of applications to GW detectors.
 The work \cite{Kohri:2018awv} provides
  relatively simple semi-analytic formulas for the spectrum ${\cal P}^{(h)}$ of  the GW induced at second-order from a scalar spectrum ${\cal P}^{(\zeta)}$.  
      ${\cal P}^{(h)}$ is written in terms of convolutions:
  \be
  {\cal P}^{(h)}_k (\tau)\,=\,\int_0^\infty\,d v\,\int_{|1-v|}^{1+v}\,d u\,
  \,{\cal K}(\tau,\,u,\,v)\,\,{\cal P}^{(\zeta)}_{u k}(\tau)
  \,{\cal P}^{(\zeta)}_{v k}(\tau)\,.
  \ee
  The expression for the kernel ${\cal K}(\tau,\,u,\,v)$  is found in \cite{Kohri:2018awv}.
 The associated energy density in GW as function of the scale is given by  (the bar
 indicates time-averaging over oscillations)
 \be
 \Omega_{\rm GW}(\tau,\,k)
 \,=\,\frac{1}{24}\,
 \left( \frac{k}{a\,H}\right)^2\,
 {\bar{\cal P}}^{(h)}_k (\tau)\,.
 \ee

\begin{figure}[h!]
\centering
 \includegraphics[width = 0.6 \textwidth]{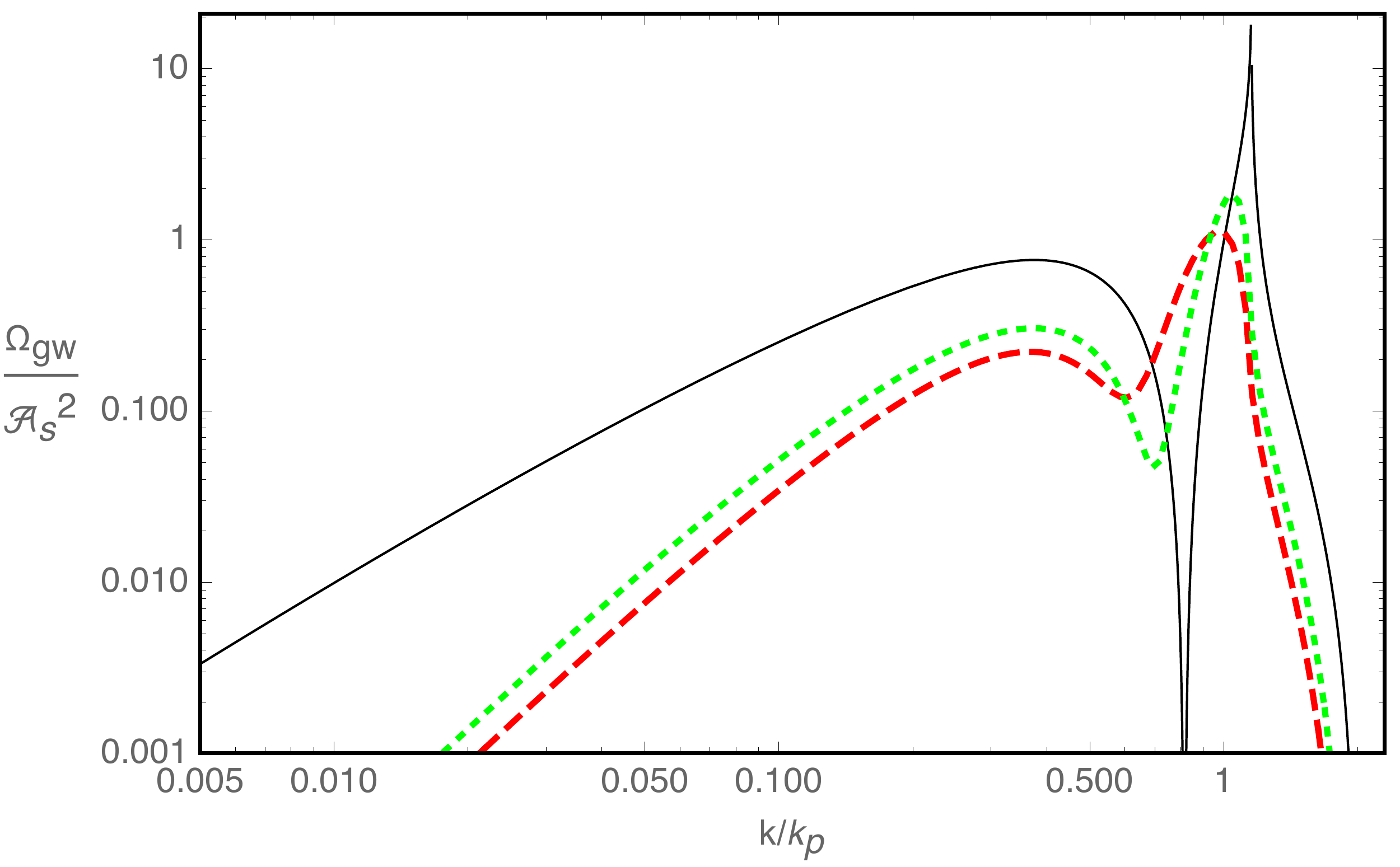}
 \caption{\it Profile of the gravitational wave  energy density $\Omega_{\rm GW}$ as a function of the scale. 
 {\it Continuous black line}:
 $\delta$-like spectrum \eqref{PSn1}. {\it Red dashed line}: $k^4$-spectrum  \eqref{PSn2}. 
 {\it Green dotted line}: $k^8$-spectrum  \eqref{PSn3}.}
 \label{fig:complete-profile3aGW}
\end{figure}

 As representative examples, we consider three cases of curvature spectrum:
 \bea
 {\cal P}^{(\zeta)}_{ k}&=&{\cal A}_s\,\delta\left(\ln{k/k_\star}\right)\,,
 \label{PSn1}
 \\
  {\cal P}^{(\zeta)}_{ k}&=&4\,{\cal A}_s\,\left( k/k_\star\right)^4
   \label{PSn2}\,,
   \\
  {\cal P}^{(\zeta)}_{ k}&=&8\,{\cal A}_s\,\left( k/k_\star\right)^8
   \label{PSn3}\,,
 \eea
 so to compare with the results of \cite{Ananda:2006af,Byrnes:2018txb}. The normalizations have been
 chosen so to ensure that $\int d \ln k\, {\cal P}^{(\zeta)}_{ k}\,=\,{\cal A}_s$ in all three cases.
 Power spectrum \eqref{PSn1} corresponds to a $\delta$-like spectrum in log-space, peaked
 at a characteristic frequency $k_\star$; the power spectra \eqref{PSn2} and \eqref{PSn3}
 correspond respectively to power-laws with the maximal slope $n-1=4$ found  in
 \cite{Byrnes:2018txb}, and the maximal slope $n-1=8$ discussed above.
 Fig \ref{fig:complete-profile3aGW} 
 represents the corresponding profiles of 
 $ \Omega_{\rm GW}(\tau,\,k)$ in each of the three cases.

\smallskip
We notice that the profile of the GW spectra in red and green are very similar, with the green one -- corresponding
to a profile for curvature perturbations spectrum scaling as $k^8$ -- slightly more amplified, and with more enhanced  dip at intermediate scales. It would be interesting to study whether future GW experiments, for example using binning techniques as proposed \cite{Caprini:2019pxz}, can distinguish among these frequency profiles of GW spectra, in case of a detection of stochastic GW background.

\section{Conclusions}
\label{sec-concl}
We developed an analytical, model independent approach to study the spectrum  of cosmological fluctuations (scalar or tensor)
 in single-field models of inflation that include brief epochs of violation of slow-roll conditions. 
 Our analysis did not make any hypothesis on the value
 of the slow-roll parameters  -- instead we implemented a gradient expansion for solving the equations of fluctuations. 
 We solved exactly 
 these equations   at leading order  in a small parameter that controls the 
 duration of non-slow-roll evolution.  Our approach leads to simple expressions for the fluctuation mode functions,
 depending on few parameters related with the slope of the pump field $z(\tau)$ during the phases of non-slow-roll evolution,  and with the  duration of the non-slow-roll eras. These parameters control the total enhancement and the slope of the fluctuation
  spectrum from large towards small scales.  
  Our  formulas for the power spectrum lead to qualitative features that are
in overall agreement 
with what found in other numerical and analytical studies of concrete inflationary models with non-slow-roll phases, 
as the presence of dips, and of phases of rapid growth in the spectrum. This fact
suggests that those features can be well described by our approximations of very short duration of non-slow-roll phase. As a byproduct of our methods, we first
proposed new quantitative relations connecting the position  of the most pronounced dip $k_{\rm dip}/k_\star$,  with the maximal size of  enhancement of the  spectrum $\Pi_{\rm max}$: $k_{\rm dip}/k_\star\,\propto \,\Pi_{\rm max}^{-1/4}$. Then, we confirmed that in the case of a single short phase of non-slow-roll evolution
the steepest growth of the spectrum has spectral index $n-1\,=\,4$; on the other hand, if multiple
  non-slow-roll phases occur then this value can increase. In an example with two non-slow-roll phases, the maximal
  spectral index is $n-1\,=\,8$,  indicating that 
   the  slope of the spectrum has memory  of the total history 
   of non-attractor phases during inflation.
    
    \smallskip
    
   These findings  can be relevant when testing against observations the predictions of single-field models of inflation  that include short non-slow-roll phases. The position of the most pronounced dip in the spectrum of fluctuations indicates the range of scales where
   predictions start to  strongly deviate from   slow-roll models. Knowledge of the dip properties
    can then be important when testing these scenarios against   constraints on the large scale spectra obtained from CMB temperature fluctuations and spectral distortions.  On the other hand, knowledge of the steepest growth of the
    spectrum is essential for understanding at what extent non-slow-roll single-field models are able to produce
    primordial black holes, or secondary stochastic gravitational wave spectra testable with current or future observations. 
    
\smallskip
Our method is based on a truncation of a Taylor series in the duration of non-slow-roll era at the first non-vanishing contribution: it is
apparently better suited to study systems where slow-roll violations happen suddenly and last briefly. On the other
hand, the resulting profiles of the fluctuation spectrum are consistent with previous literature in which the non-slow-roll phases can be more prolonged. It would be very interesting to understand more completely  the relations 
between different methods and findings. Moreover, in this work we focussed on 2-point correlation
functions in single-field inflation: it would be interesting to apply our methods also to higher-point correlation functions, or to models of multiple field inflation with epochs of slow-roll violation. We plan to develop these topics
in future publications.

\subsection*{Acknowledgments}

It is a pleasure to thank Debika Chowdhury, Ogan \"Ozsoy and Ivonne Zavala for useful discussions, and input on  
the manuscript.  
GT is partially funded by STFC grant ST/P00055X/1.

\begin{appendix}
\section{An example with arbitrarily fast decreasing  pump field}\label{App_toymodel}

In this Appendix, we identify the mode function $\psi_k$ with the curvature perturbation $\zeta_k$
in single field inflation with canonical kinetic terms.  
 For simplicity, we work here with coordinate time $t$ related to conformal time by $d \tau\,\equiv\,d t/a(t)$. 
Calling $\phi$ the homogeneous profile the scalar
inflaton,
  the
pump field
reads
\be\label{pmpsfi}
z(t)\,=\,a(t)\,\frac{\dot \phi(t) }{H(t)}\,.
\ee
If $\dot \phi$ and $H$ are nearly constant as in single field slow-roll inflation, then 
the pump field $z\propto a$;  so it increases
 with time during inflation (in conformal time, its profile is well approximated by eq. \eqref{srpf}). 
 On the other hand,  models exist where the scalar velocity $\dot \phi $ transiently decrease very fast, even in a regime of nearly 
 de Sitter expansion. In those cases, the pump field decreases with time, and we enter in a brief regime of non-attractor.

 In this Appendix we build a representative single-field model to demonstrate that the   decreasing rate can be arbitrarily large. Working in nearly de Sitter space, 
 we consider  a system where during a short epoch the scalar derivative
 decreases arbitrarily fast. We build the model following the approach of Starobinsky \cite{Starobinsky:1992ts}, who considered cases of a linear potential $V(\phi)$ that is continuous, but with a sudden change of slope at a certain value $\bar \phi$ of the scalar. In such a system, for a short time interval, the scalar velocity $\dot \phi$ decreases as the cube of the scale factor, $1/a^3$, and during such interval  
 the pump field decrease with time.

\smallskip

We now show that, by selecting an appropriate set-up, the scalar velocity can decrease during a short non-attractor phase as 
$\dot \phi \propto 1/a^{\beta}$
 for arbitrary $\beta$.  This dimensionless parameter $\beta$ controls 
  the rate of variation of the pump field. 
For simplicity, to present our system it is convenient to adopt a Hamilton-Jacobi approach \cite{Salopek:1990jq,Byrnes:2009qy}, in which the Hubble parameter is expressed as function of the scalar: $H\,=\,H(\phi)$. Setting $M_{\rm Pl}\,=\,1$, the system of homogeneous equations reads
\bea
\dot \phi&=&-2\,H_{,\phi}(\phi) \label{eqvel1}
\,,\\
H^2(\phi)&=&\frac{V(\phi)}{3}+\frac{2\,H_{,\phi}(\phi)^2}{3} \label{eqvel1A}\,,
\eea
with $V(\phi)$ the scalar potential. It is not hard to show that the results of this method are equivalent
to the more traditional approach based on Lagrange equations, in the limit of homogeneous evolution (see \cite{Salopek:1990jq}).

\smallskip

Inspired by the Starobinsky model  \cite{Starobinsky:1992ts}, we choose a profile for the Hubble parameter that is continuous but has a 
jump in its first derivative at $\phi=\bar\phi$. It is
\be\label{Htoy1}
H(\phi)\,=\,H_0+\frac{H_1}{ 2}\,\phi +\frac{H_1\,H_2^{(\pm)}}{2 \beta\,H_0}\,e^{-\beta\,\frac{H_0}{H_1}\,\phi}
+\, \Theta[\phi-\bar \phi]\,\left( 
\frac{H_{2}^{(-)}
-H_{2}^{(+)}
}{2 \beta H_0}\,H_1\,e^{-\beta \frac{H_0}{H_1}\,\bar \phi}
\right)\,,
\ee
where $H_0$, $H_1$, $H_2$, $\beta$ are arbitrary constants. The suffixes $(\pm)$ in   $H_{2}$ indicate
that this quantity is computed on the right or on the left of $\bar \phi$.  
The constant quantity $H_0$ shall represent the nearly constant value of the Hubble parameter
during inflation, that receives small correction by the remaining terms (we assume $H_{1,2}\ll H_0$). In such case, 
during all stages $a(t)\,=\,e^{H_0 t}$ at good approximation. 
The last term in expression \eqref{Htoy1} is proportional to a Heaviside $\Theta$-function and makes
the Hubble parameter continuous at the position $\phi\,=\,\bar \phi$. On the other hand, the potential $V(\phi)$,
which can be read from eq \eqref{eqvel1A},
has discontinuities in $\phi$.

It is straightforward to solve eq. \eqref{eqvel1}. The scalar field velocity results
\be
\dot\phi(t)\,=\,-\frac{H_1^2\,\phi_0\,e^{-\beta\, H_0\,t}}{H_2^{(\pm)}+H_1\,\phi_0\,e^{-\beta H_0 t}}\,,
\ee
with $\phi_0$ an arbitrary constant. 
We start with values $\phi>\bar \phi$, and we assume that in such stage $H_2^{(+)} \ll H_1 \,\phi_0\,e^{-\beta H_0 t}$. Then, in this phase $\dot\phi\simeq -H_1$, and the pump field \eqref{pmpsfi} increases with time as in single-field
inflation.

Then, at $\phi\,=\,\bar \phi$, we enter in the second region where $H_2^{(-)}\,\gg\,H_1 \,\phi_0\,e^{-\beta H_0 t} $. In
such case, the scalar velocity decreases very rapidly
\bea
\dot \phi&\simeq&-\frac{H_1^2\,\phi_0\,e^{-\beta\, H_0\,t}}{H_2^{(-)}}\,\hskip1cm\Rightarrow\hskip1cm
\dot \phi
\,\propto\,\frac{1}{a^\beta}\,,
\eea
hence, since $\beta$ is a free parameter, the the pump field \eqref{pmpsfi} can decrease as rapidly as we desire. 
 This non-attractor phase can then end by adding yet another contribution to the Hubble parameter to slow-down
 again the scalar velocity. This concrete example  shows that there is in principle no bound on the rate
 of decrease of the pump field in single field inflation. 
 
 \section{Spectral index and multiple non-slow-roll epochs}\label{App_comparison}

In section \ref{secMNS} we found that for inflationary processes containing two eras of non-slow-roll evolution,
separated by an epoch of standard slow-roll expansion, the slope of the spectrum can increase from $n-1\,=\,4$
to $n-1\,=\,8$  well after the first dip. 
 Here we study the same phenomenon using  two alternative approaches, to qualitatively confirm
 our findings. We focus on the dynamics of the curvature perturbation $\zeta_k$ in single-field
 inflation, for which the pump
 field is given by \eqref{pmpsfi}.
 
 \subsubsection*{Israel junction conditions}

 The first approach  we use is the analytical
   method of
  Israel junction conditions between phases of constant slow-roll parameters, as used in
    \cite{Byrnes:2018txb}.

\begin{figure}[h!]
\centering
 \includegraphics[width = 0.49 \textwidth]{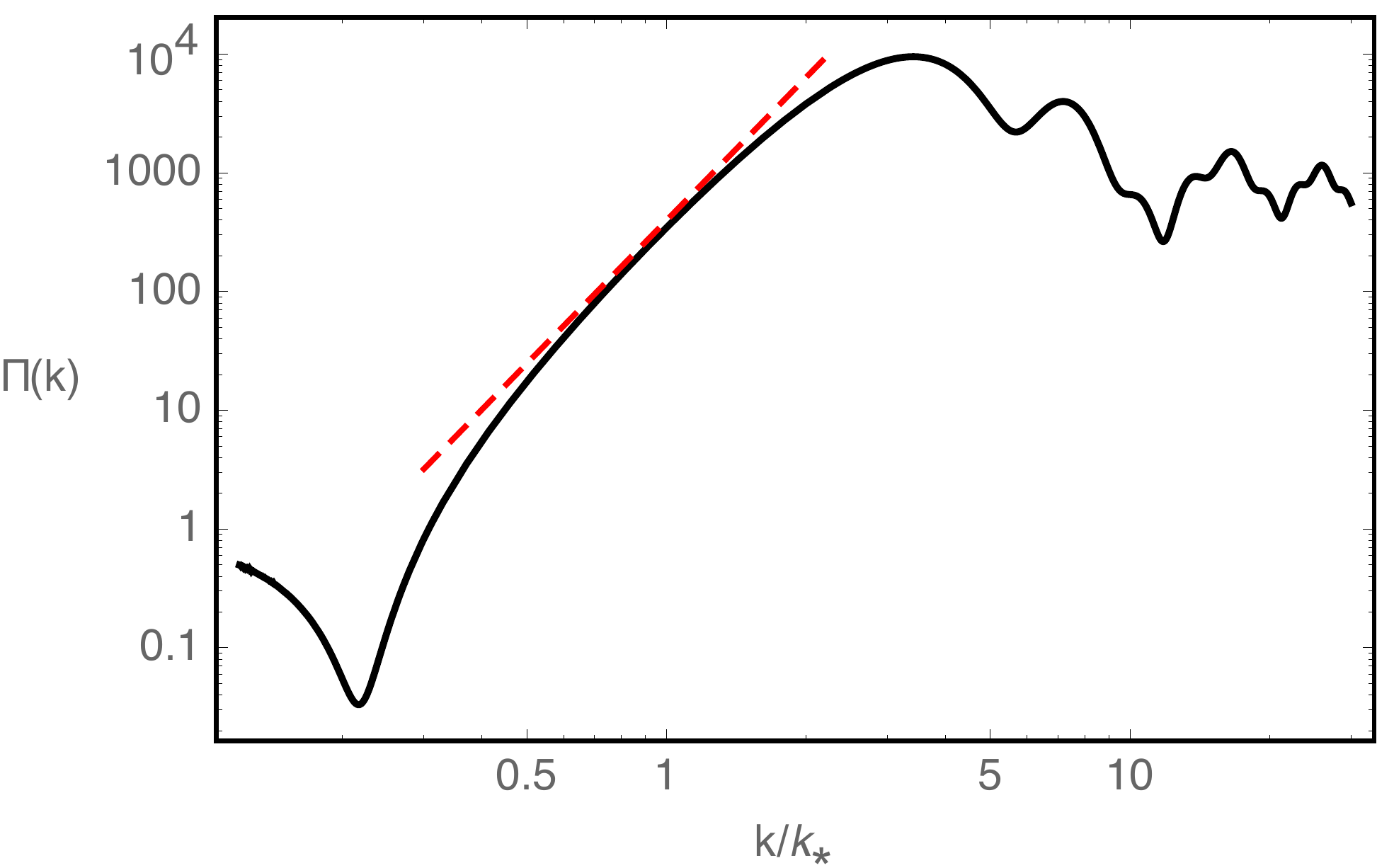}
  \includegraphics[width = 0.49 \textwidth]{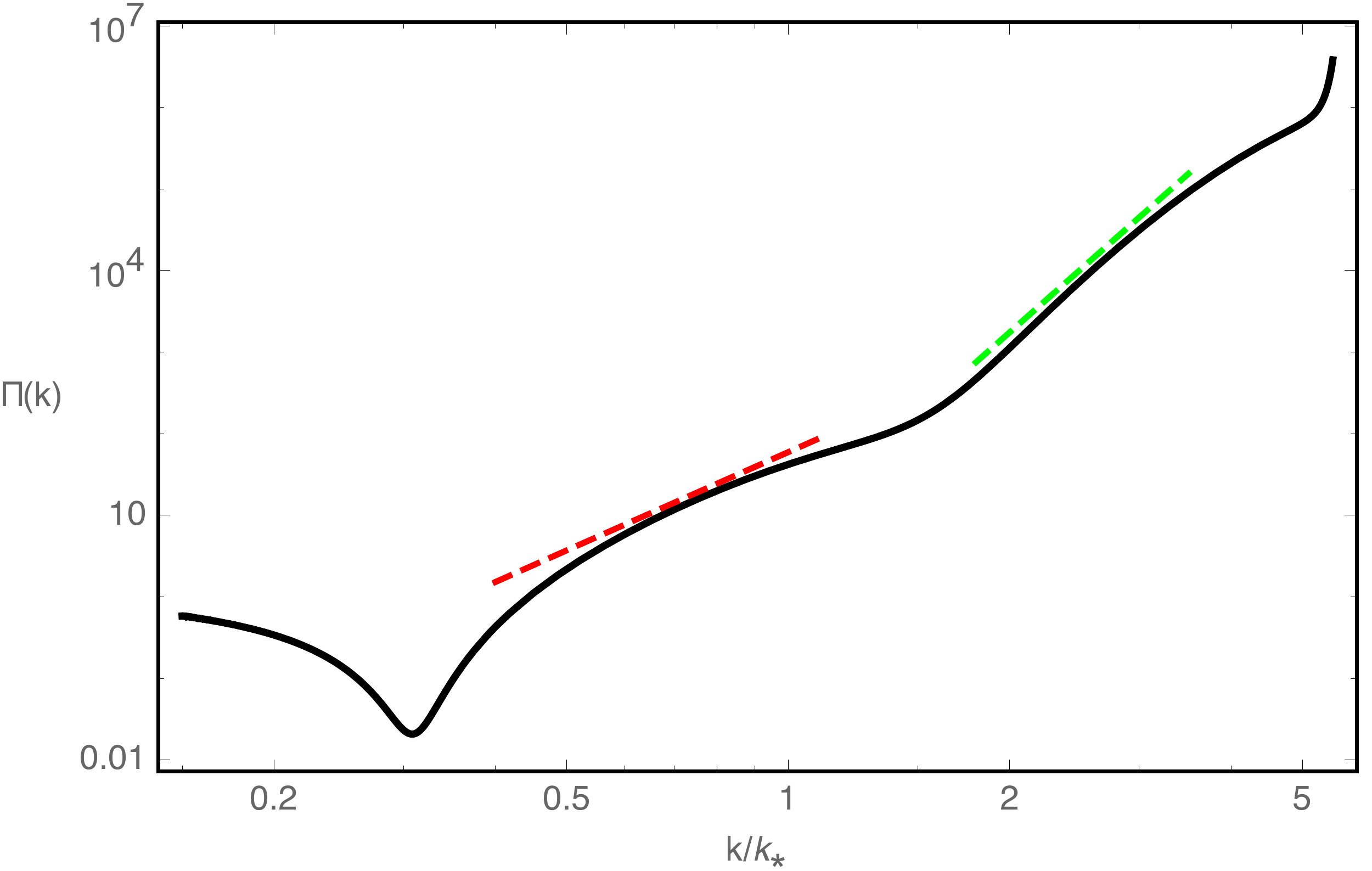}
 \caption{\it 
  Spectrum of super-horizon fluctuations evaluated at the end of inflation for an inflationary system containing one (left panel) or two (right panel)
  phases of ultra-slow-roll inflation. This plot was produced
  using a method based on Israel junction conditions, as  \cite{Byrnes:2018txb}. Red dashed lines have slope $k^4$,
  green dashed lines have slope $k^8$. 
 {\bf Left} A single-phase of non-attractor. {\bf Right} two phases of non-attractor, lasting respectively $0.2$ e-fold and $1.7$ e-fold, containing a $1.3$ e-fold epoch of normal slow-roll in between.}
 \label{fig:complete-profile3aGW1}
\end{figure}

  We study the simplest possible situation, where two phases of ultra-slow-roll evolution characterized by slow-roll parameters $\epsilon_1\,\simeq\,0$ and $\epsilon_2\,\simeq\,-6$ (see the definition around eq \eqref{defsr1}) contain in between a phase
of quasi-de Sitter expansion (where for simplicity, we set $\epsilon_1\,\simeq\,\epsilon_2\,=\,0$).  The system
is pictorially depicted in Fig \ref{fig:arte2}, where the green regions are the slow-roll, the blu regions the ultra-slow-roll phases.

 The general mode functions during a  pure de Sitter evolution are given in eq \eqref{phidsG}. During ultra-slow-roll, 
a straightforward computation (or the duality arguments of \cite{Wands:1998yp}) gives the following expression
for the mode function $\psi_k^{\rm USR}(\tau)$ solving the evolution equations
\be\label{phiUSG}
\psi_k^{\rm USR}(\tau)\,=\,\left( \frac{\bar \tau}{\tau}\right)^3\,\psi_k^{\rm dS}(\tau)\,,
\ee
where $\bar \tau$ is a reference time scale, and $\psi_k^{\rm dS}(\tau)$ the general solution \eqref{phidsG}
during the de Sitter epoch. Applying standard Israel matching conditions at the boundaries between phases as
in  \cite{Byrnes:2018txb}, we get the expression of 
mode functions that allows one to compute the spectrum of super-horizon modes at the end of inflation.

Fig
\eqref{fig:complete-profile3aGW1} (left panel) shows the resulting spectrum after a single phase of non-attractor.
Within this
approach based on Israel conditions 
 the position of the dip and its relation to the maximal growth of the spectrum can be analytically computed. In fact,
we 
already shown in the main text -- see our discussion around eq \eqref{dimcoby}, comparing our results  with   \cite{Byrnes:2018txb}  --  that  relation \eqref{kdipapp}, stating that  $k/k_\star\propto \sqrt{3/2} \,\Pi_{\rm max}^{-1/4}$, is obeyed fairly well.

 Fig
\eqref{fig:complete-profile3aGW1} (right panel) contains an example of spectrum evaluated following these methods,  which clearly shows that an initial grow with (at most) $n_S-1\,=\,4$ spectral tilt is followed by a second phase with
where the spectral tilt can reach $n_S-1\,=\,8$.

\subsubsection*{Numerical method}

We can investigate this phenomenon using numerical methods applied to a simple generalization of Starobinsky model \cite{Starobinsky:1992ts}.

\begin{figure}[h!]
\centering
 \includegraphics[width = 0.49 \textwidth]{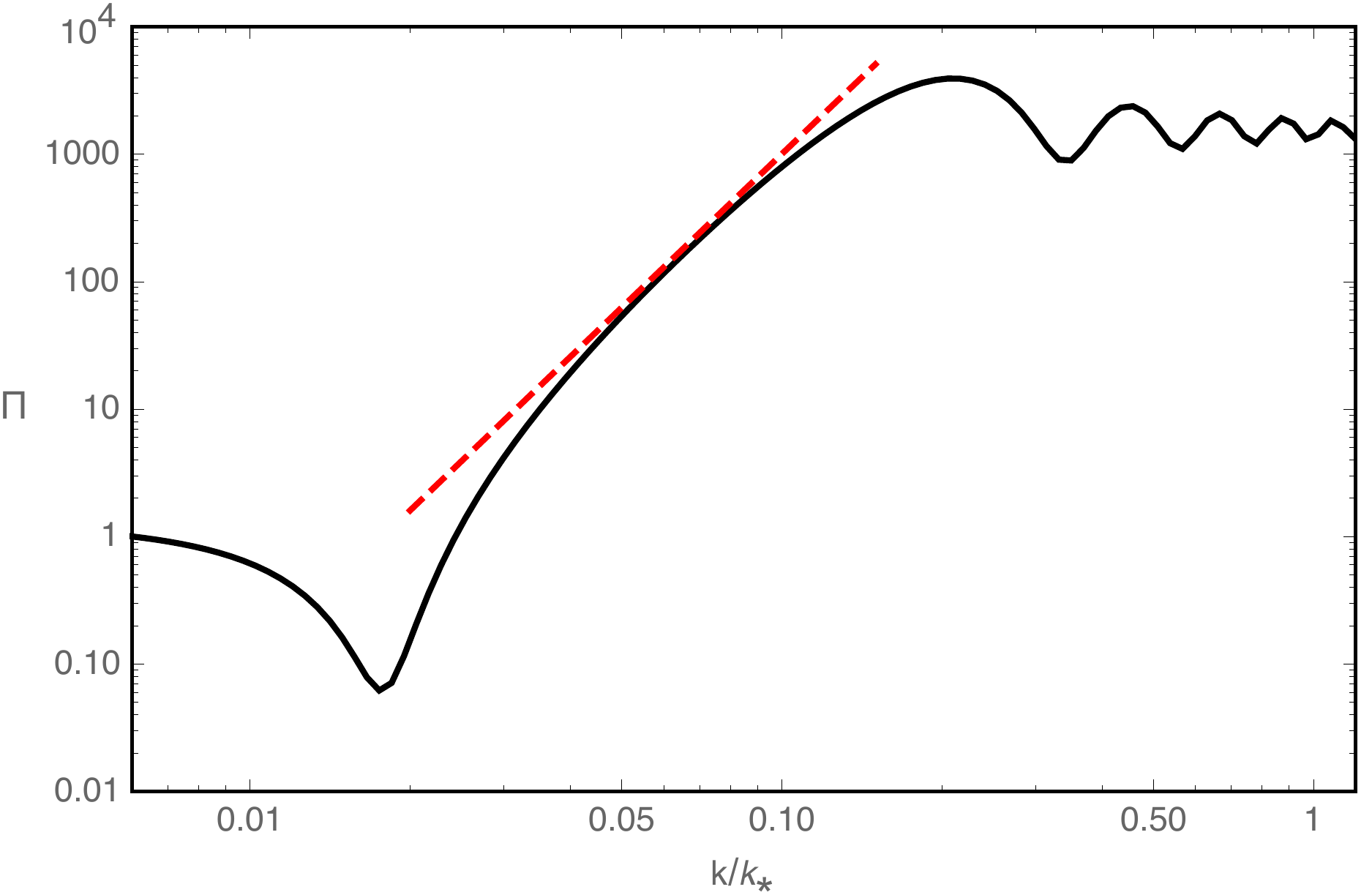}
  \includegraphics[width = 0.49 \textwidth]{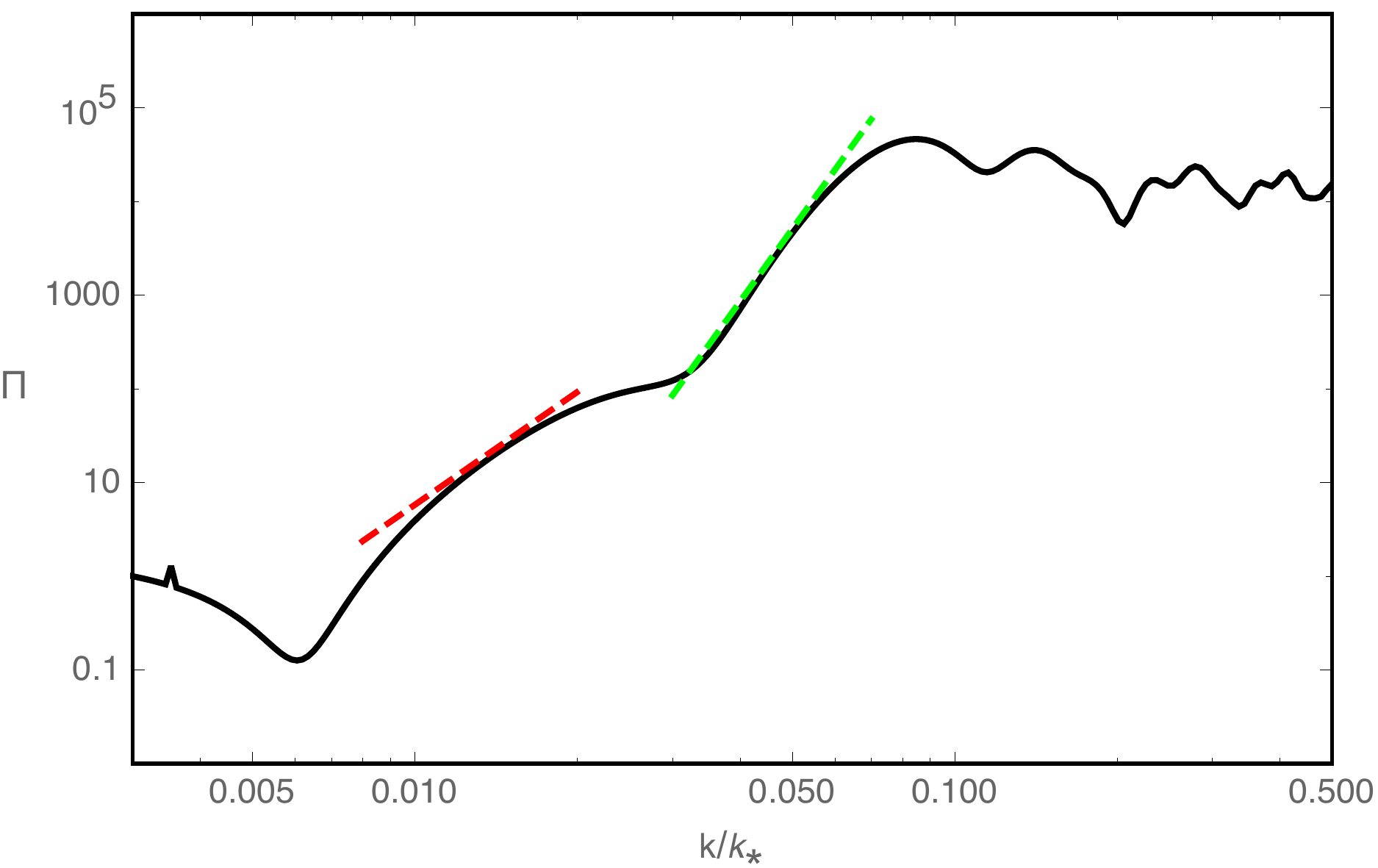}
 \caption{\it 
  Spectrum of super-horizon fluctuations 
  evaluated at the end of inflation for an inflationary system containing one (left panel), or two (right panel)
  phases of non-attractor  evolution. This plot was produced
  starting
  from the generalization \eqref{potstarg} of the Starobinsky model \cite{Starobinsky:1992ts,Leach:2001zf}.
   Red dashed lines have slope $k^4$,
  green dashed lines have slope $k^8$. 
 {\bf Left} A single-phase of non-attractor. {\bf Right} two phases of non-attractor 
 with a nested epoch of standard slow-roll expansion. The first non-attractor phase lasts $1.4$ e-fold, and the duration of slow-roll expansion is $2$ e-folds.
 We choose the parameters $A_1=90$, $A_2=15$, $A_4=3$, while $A_3$ satisfies  conditions to ensure continuity of the scalar first derivative.}
 \label{fig:complete-profile3aGW2}
\end{figure}

 The model of \cite{Starobinsky:1992ts} (see also \cite{Leach:2001zf}, the discussion in their section IV) consists
of a linear inflationary potential $V(\phi)$ that is continuous, but with a jump in its first derivative at a given value of the scalar $\phi$. We can generalize the potential allowing for several discontinuities in the first derivative of the potential:
\be \label{potstarg}
V(\phi)\,=\,\left\{
    \begin{array}{rl}
     V_0+A_1 \left( \phi-\phi_1\right)\, & \hskip0.5cm \text{for } \phi\, >\, \phi_1,\\
     V_0+A_2 \left( \phi-\phi_1\right)\, & \hskip0.5cm \text{for } \phi_2\le \phi\, \le\, \phi_1,\\
     V_0+A_3 \left( \phi-\phi_2\right)\, & \hskip0.5cm \text{for } \phi_3\le \phi\, \le\, \phi_2,\\
     V_0+A_4 \left( \phi-\phi_3\right)\, & \hskip0.5cm \text{for } \phi\, <\, \phi_3\,.\\
    \end{array} \right.
\ee
where $V_0$, $A_i$, $\phi_i$ are constant parameters. We choose $V_0$ large enough to ensure
that the scale factor can be well approximated by de Sitter space with a nearly constant Hubble
parameter $H_0$ and $a(t)\simeq e^{H_0 t}$.  It is simple to solve the evolution equation for the scalar field $\phi$ in each of the ranges appearing in eq \eqref{potstarg}. The structure of the solution is
\be
3 H_0\,\dot \phi(t)\,=\,-A_i+c_i/a^3(t)
\ee
where the $A_i$ are the parameters appearing in eq \eqref{potstarg}, while the $c_i$ are integration constants
chosen to ensure that the first derivative is continuous at the transition points $\phi_i$. 
As shown in  \cite{Starobinsky:1992ts},  \cite{Leach:2001zf}, 
if there is a large hierarchy between the quantities $A_i$, one can design situations where periods of slow-roll
evolution with $3 H_0 \dot  \phi\simeq$ constant are nested between periods of non-attractor phases with 
$3 H_0 \dot  \phi\,\propto\,1/a^3(t)$.  Figure \ref{fig:complete-profile3aGW2} contains two examples of 
 numerical evaluation of the curvature fluctuation spectrum for this
 system.

  In the left we plot the original Starobinsky example \cite{Starobinsky:1992ts,Leach:2001zf}, with a single discontinuity in the first derivative of the potential, leading to a single phase of non-attractor evolution: the spectrum of curvature fluctuations grow with a maximal slope $k^4$.  In the original example the single non-attractor phase is relatively prolonged, in order to sufficiently increase the spectrum. Hence we can not expect that our formula
   \eqref{kdipapp} relating dip position with the spectrum maximal enhancement is too accurate, 
   since we derived eq   \eqref{kdipapp}  in the limit of short non-attractor phase. On the other hand, we numerically evaluated the ratio of   the dip position  versus the scale of non-attractor inception for our example, finding the value
    $0.27$. By computing the value of maximal enhancement  of the spectrum we find the relation  $\sqrt{3/2} \,\Pi_{\rm max}^{-1/4}\,=\,0.19$. Hence, eq  \eqref{kdipapp}  is satisfied up to a factor of order $1.5$.

  In the right plot of figure \ref{fig:complete-profile3aGW2} we instead use the complete potential \eqref{potstarg}, and we study
two phases of non-attractor, with a nested epoch of slow-roll in between. We notice that towards small
scales the spectrum increases its slope reaching $n_S-1\,=\,8$, as desired.

 \bigskip

These preliminary findings support the results of the main text.  It would be interesting to carefully study the system in a more realistic set-up. We defer this question to future work.

\end{appendix}







\addcontentsline{toc}{section}{References}
\bibliographystyle{utphys}

\mciteSetMidEndSepPunct{}{\ifmciteBstWouldAddEndPunct.\else\fi}{\relax}

\providecommand{\href}[2]{#2}\begingroup\raggedright\endgroup


\end{document}